\documentclass[aps,prxquantum,reprint,twocolumn,superscriptaddress,floatfix,nofootinbib,longbibliography]{revtex4-2}
\usepackage{graphicx,amsmath,amsfonts,amssymb,amsthm,xr}
\usepackage{epsfig,amsmath,amssymb,color,dsfont,upgreek,physics}
\usepackage{mathrsfs}
\usepackage{mathtools}
\usepackage{bbold}
\usepackage{float}
\usepackage[caption=false]{subfig}

\usepackage[bookmarks=true,colorlinks,linkcolor=OrangeRed,urlcolor=NavyBlue,citecolor=RoyalBlue]{hyperref}
\usepackage[dvipsnames]{xcolor}
\usepackage{orcidlink}

\usepackage{graphicx}
\usepackage{dcolumn}
\usepackage{bm}
\usepackage{color}

\definecolor{mypurple}{rgb}{0.49,0.18,0.56}

\definecolor{limeGreen}{rgb}{0.55, 0.71, 0.0}

\usetikzlibrary{arrows.meta}
\usepackage{changes}
\usetikzlibrary{quantikz2}
\usepackage{helvet}
\usepackage[tracking=true]{microtype}
\SetTracking{encoding=*}{90} 

\begin{document}
\title{Efficient Qudit Circuit for Quench Dynamics of $2+1$D Quantum Link Electrodynamics}

\author{Rohan Joshi}
\affiliation{Max Planck Institute of Quantum Optics, 85748 Garching, Germany}
\affiliation{Munich Center for Quantum Science and Technology (MCQST), 80799 Munich, Germany}

\author{Michael Meth${}^{\orcidlink{0000-0002-5446-3962}}$}
\affiliation{Universit\"at Innsbruck, Institut f\"ur Experimentalphysik, 6020 Innsbruck, Austria}

\author{Jan C.~Louw${}^{\orcidlink{0000-0002-5111-840X}}$}
\affiliation{Max Planck Institute of Quantum Optics, 85748 Garching, Germany}
\affiliation{Munich Center for Quantum Science and Technology (MCQST), 80799 Munich, Germany}

\author{Jesse J.~Osborne${}^{\orcidlink{0000-0003-0415-0690}}$}
\affiliation{Max Planck Institute of Quantum Optics, 85748 Garching, Germany}
\affiliation{Munich Center for Quantum Science and Technology (MCQST), 80799 Munich, Germany}
\author{Kevin Mato}
\affiliation{Chair for Design Automation, Technical University of Munich, Munich, Germany}
\author{Martin Ringbauer${}^{\orcidlink{0000-0001-5055-6240}}$}
\email{martin.ringbauer@uibk.ac.at}
\affiliation{Universit\"at Innsbruck, Institut f\"ur Experimentalphysik, 6020 Innsbruck, Austria}
\author{Jad C.~Halimeh${}^{\orcidlink{0000-0002-0659-7990}}$}
\email{jad.halimeh@physik.lmu.de}
\affiliation{Max Planck Institute of Quantum Optics, 85748 Garching, Germany}
\affiliation{Department of Physics and Arnold Sommerfeld Center for Theoretical Physics (ASC), Ludwig Maximilian University of Munich, 80333 Munich, Germany}
\affiliation{Munich Center for Quantum Science and Technology (MCQST), 80799 Munich, Germany}

\begin{abstract}
A major challenge in the burgeoning field of quantum simulation for high-energy physics is the realization of scalable $2+1$D lattice gauge theories on state-of-the-art quantum hardware, which is an essential step towards the overarching goal of probing $3+1$D quantum chromodynamics on a quantum computer. Despite great progress, current experimental implementations of $2+1$D lattice gauge theories are mostly restricted to relatively small system sizes and two-level representations of the gauge and electric fields. Here, we propose a resource-efficient method for quantum simulating $2+1$D spin-$S$ $\mathrm{U}(1)$ quantum link lattice gauge theories with dynamical matter using qudit-based quantum processors. By integrating out the matter fields through Gauss's law, we reformulate the quantum link model in a purely spin picture compatible with qudit encoding across arbitrary spatial dimensions, eliminating the need for ancillary qubits and reducing resource overhead. Focusing first on the spin-$1/2$ case, we construct explicit circuits for the full Hamiltonian and demonstrate through numerical simulations that the first-order Trotterized circuits accurately capture the quench dynamics even in the presence of realistic noise levels. Additionally, we introduce a general method for constructing coupling-term circuits for higher-spin representations $S>1/2$. Compared to conventional qubit encodings, our framework significantly reduces the number of quantum resources and gate count. Our approach significantly enhances scalability and fidelity for probing nonequilibrium phenomena in higher-dimensional lattice gauge theories, and is readily amenable to implementation on state-of-the-art qudit platforms.
\end{abstract}

\maketitle
\tableofcontents

\section{Introduction}
Lattice gauge theories (LGTs) \cite{Kogut1975,Rothe_book} are a powerful framework originally developed in high-energy physics (HEP) to shed light on the problem of quark confinement in quantum chromodynamics (QCD) \cite{Wilson1974}. Defined on a discretized lattice, they describe the gauge-invariant coupling of matter to electric and gauge fields. In the continuum limit, they give rise to gauge theories, which lie at the heart of the Standard Model of particle physics in describing interactions between elementary particles as mediated through gauge bosons \cite{Weinberg_book,Zee_book,Peskin2016}. 

Through Monte Carlo techniques, especially in the Euclidean path integral formulation, LGTs have enabled precise computation of hadron masses, decay constants, and thermodynamic properties of QCD, to name a few \cite{Creutz1980MC,QCD_review,Gattringer_book,Montvay_book}. Very quickly, LGTs proved to be equally powerful tools in other fields as well. In condensed matter, LGTs serve as models of emergent gauge structures in frustrated magnets and quantum spin liquids \cite{Wegner1971,Kogut_review,wen2004quantum,Savary2016,Calzetta_book}. In quantum many-body physics, they are an important platform for various nonergodic phenomena and intriguing properties uncovered through tensor network \cite{Uli_review,Orus2013,Paeckel_review,Orus2019,Montangero_book} and exact diagonalization techniques \cite{Sandvik2010}, including quantum many-body scarring \cite{Surace2020,Desaules2022weak,Desaules2022prominent,aramthottil2022scar}, disorder-free localization \cite{Smith2017,Brenes2018,smith2017absence,karpov2021disorder,Sous2021,Chakraborty2022,Halimeh2021enhancing}, nonstabilizerness or magic \cite{Tarabunga2023many,hartse2024stabilizerscars,Smith2025nonstabilizerness,Falcao2025Nonstabilizerness,Esposito2025magic}, and Hilbert-space fragmentation \cite{Desaules2024ergodicitybreaking,desaules2024massassistedlocaldeconfinementconfined,jeyaretnam2025hilbertspacefragmentationorigin,ciavarella2025generichilbertspacefragmentation}.

Despite the above tremendous progress, there are still major challenges when it comes to studying LGTs. Monte Carlo techniques are not suited to large matter densities and out-of-equilibrium dynamics due to the sign problem \cite{deforcrand2010simulatingqcdfinitedensity,Troyer2005computational}. Tensor network techniques suffer in higher spatial dimensions and are limited in maximally accessible evolution times due to the exponential growth of bond dimension \cite{Banuls_review,magnifico2024tensornetworkslatticegauge}. As such, an alternative venue that can offer an \textit{ab initio} approach to uncovering the out-of-equilibrium properties of LGTs in $d+1$D ($d$ spatial and one temporal dimensions) is extremely timely.

Quantum simulation for high-energy physics (HEP) has seen remarkable progress in recent years \cite{Dalmonte_review, Zohar_review, Aidelsburger2011, Zohar_NewReview, klco2021standard,Bauer_ShortReview, Bauer_review, dimeglio2023quantum, Cheng_review, Halimeh_review, Cohen:2021imf, Lee:2024jnt, Turro:2024pxu,bauer2025efficientusequantumcomputers}. Quantum simulators offer an alternative venue to probe HEP phenomena from first-principles time evolution, with the attractive possibility of capturing snapshots of highly nonperturbative intermediate-time dynamics that is hard to capture in dedicated particle colliders \cite{Bauer_review,dimeglio2023quantum}. The quantum advantage offered by quantum simulators renders the required resources polynomial in system size, thus making such platforms highly advantageous for simulating out-of-equilibrium dynamics in higher spatial dimensions compared with the exponential cost of classical methods.

To facilitate the experimental implementation of LGTs, the electric and gauge degrees of freedom, which are locally infinite-dimensional in the continuum limit, are truncated down to a finite-dimensional local Hilbert space using the quantum link model (QLM) formulation \cite{Chandrasekharan1997,Wiese_review,Kasper2017}. A further simplification can be made by utilizing Gauss's law to integrate out the matter degrees of freedom, which is possible in any spatial dimension $d$ in the case of bosonic matter \cite{Surace2020}. An alternative to the QLM formulation involves using Gauss's law to integrate out the gauge fields, which is generally possible only in one spatial dimension \cite{Muschik2017,Atas:2021ext}. This leads to an LGT Hamiltonian with only matter degrees of freedom, albeit with exotic infinite-range interactions, which limits the size of implementable systems. These and other schemes have allowed the experimental realization of different LGTs on various quantum hardware platforms where different fascinating phenomena have been quantum simulated \cite{Martinez2016,Klco2018,Goerg2019,Schweizer2019,Mil2020,Yang2020,Wang2021,Su2022,Zhou2022,Wang2023,Zhang2023,Ciavarella2024quantum,Ciavarella:2024lsp,Farrell:2023fgd,Farrell:2024fit,zhu2024probingfalsevacuumdecay,Ciavarella:2021nmj,Ciavarella:2023mfc,Ciavarella:2021lel,Gustafson:2023kvd,Gustafson:2024kym,Lamm:2024jnl,Farrell:2022wyt,Farrell:2022vyh,Li:2024lrl,Zemlevskiy:2024vxt,Lewis:2019wfx,Atas:2021ext,ARahman:2022tkr,Atas:2022dqm,Mendicelli:2022ntz,Kavaki:2024ijd,Than:2024zaj,Angelides2025first,cochran2024visualizingdynamicschargesstrings,gyawali2024observationdisorderfreelocalizationefficient,gonzalezcuadra2024observationstringbreaking2,crippa2024analysisconfinementstring2,schuhmacher2025observationhadronscatteringlattice,davoudi2025quantumcomputationhadronscattering}.
\begin{figure*}[t] 
    \centering
    \captionsetup[subfloat]{position=top, oneside, margin={-0.25cm,0cm}}
    \subfloat[\vspace{-0.3cm}]{%
        \hspace*{0.17cm}\includegraphics[width=1.03\textwidth]{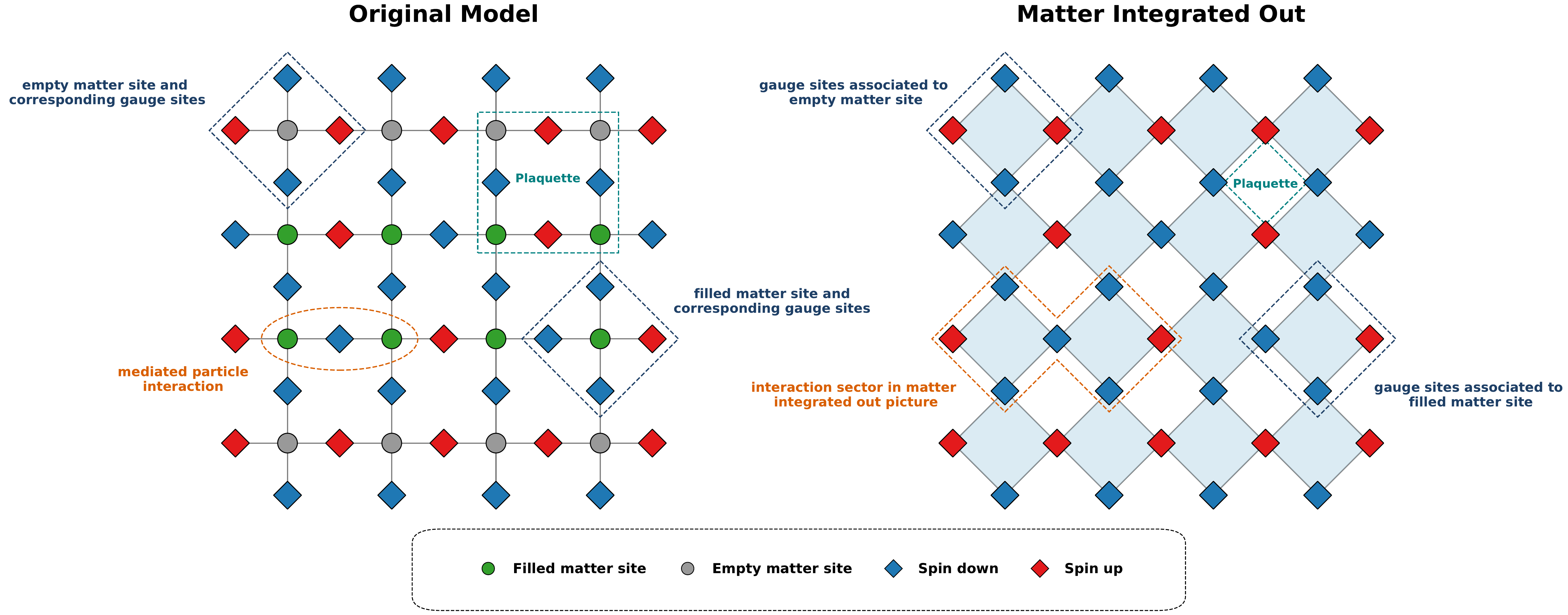}
    }
    \hfill
\captionsetup[subfloat]{
margin={-2.03cm,0cm}}

\subfloat[\footnotesize \sffamily\bfseries\textls{Gauge-invariant states}]{

        \includegraphics[width=0.8\textwidth]{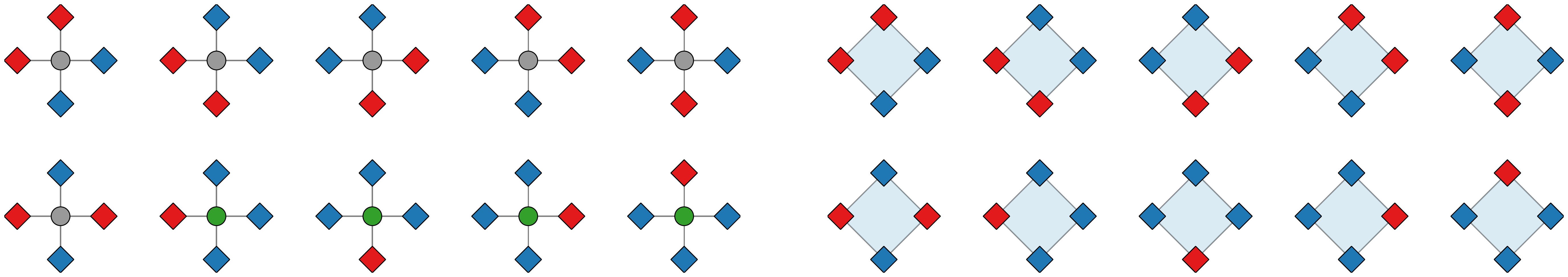}%
    }%
    \hfill
\captionsetup[subfloat]{oneside,margin={-2.52cm,0cm}}
\subfloat[\footnotesize\sffamily\bfseries\textls{Initial states}]{%
  \hspace*{-1cm}%
\includegraphics[width=0.8\textwidth]{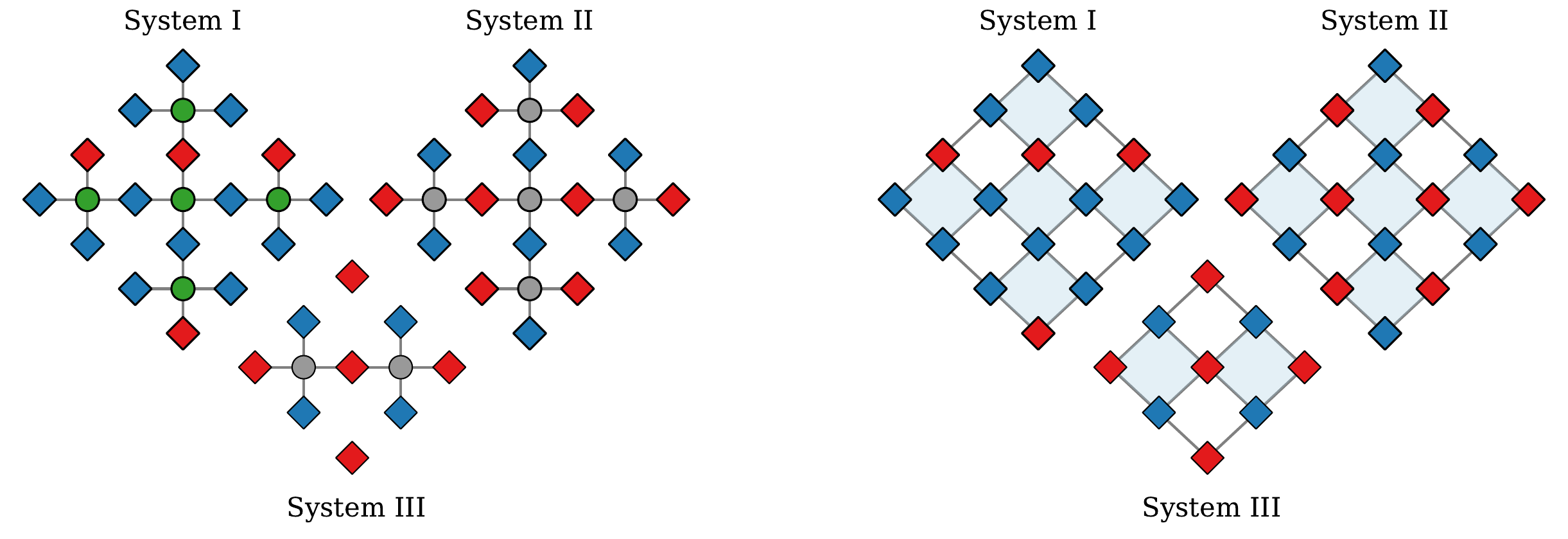}%
}\hfill
\captionsetup[subfloat]{oneside,margin={-3.75cm,0cm}}
\subfloat[\footnotesize\sffamily\bfseries\textls{Qudit encoding}]{%
\vspace*{-5cm}
\includegraphics[width=0.6\textwidth]{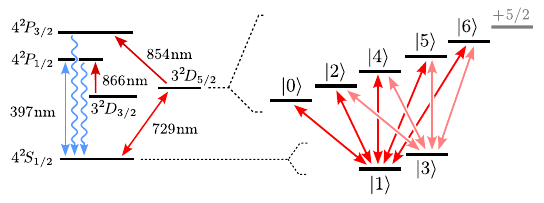}%
}

    \caption{(a) 2D lattice illustrating interactions in both original and matter-integrated-out (MIO) formulations, (b) configurations satisfying Gauss's law, (c) depicts the initial states considered in our simulations, and (d) shows a possible encoding of a qudit in the electronic states of a $^{40}\text{Ca}^+$ ion. The basis states $\ket{i}_{i = 0,\dotsc 6}$ are  encoded in the $4^2S_{1/2}$ ground states and the metastable $3^2D_{5/2}$ excited states, which offer a natural lifetime of $1.17~\text{s}$. State manipulation is realized by coherent laser pulses driving narrow optical quadrupole transitions (arrows). Short-lived dipole transitions (left) enable efficient laser cooling and state readout by fluorescence detection.}
    \label{fig:lattice_scheme}
\end{figure*}

However, challenges abound when it comes to (i) how to efficiently encode gauge symmetry in a quantum simulator in higher spatial dimensions \cite{cochran2024visualizingdynamicschargesstrings,gyawali2024observationdisorderfreelocalizationefficient,gonzalezcuadra2024observationstringbreaking2,crippa2024analysisconfinementstring2}, (ii) how to control gauge symmetry breaking errors \cite{Halimeh2020a,Halimeh2020e,Halimeh_BriefReview}, (iii) how to achieve the continuum limit of the underlying gauge theory \cite{Buyens2014,Zache2021achieving,Halimeh2021achieving,kane2025obtainingcontinuumphysicsdynamical}, and (iv) how to make designs scalable such that, when fault tolerance \cite{Acharya2025QEC,bluvstein2025architecturalmechanismsuniversalfaulttolerant,sommers2025observationfaulttolerancethreshold,dasu2025breakingmagicdemonstrationhighfidelity} is achieved, HEP quantum simulators will become \textit{bona fide} complementary venues for particle colliders. At the frontier in the field currently is the realization of $2+1$D LGTs on quantum hardware \cite{cochran2024visualizingdynamicschargesstrings,gyawali2024observationdisorderfreelocalizationefficient,gonzalezcuadra2024observationstringbreaking2,crippa2024analysisconfinementstring2}, which encompasses all the aforementioned challenges. Of particular interest in that regard is the implementation of $2+1$D lattice quantum electrodynamics (QED) on a quantum simulator and the investigation of its equilibrium and dynamic properties. There have been several proposals on how to do this in a pure LGT setting (without dynamical matter) \cite{Buechler2005,Zohar2011,Tagliacozzo2013,Dutta2017,Ott2020scalable,Fontana2022} as well as those with dynamical matter \cite{Zohar2013,Paulson2020simulating}. Whereas Ref.~\cite{Paulson2020simulating} focused on a variational quantum circuit to compute equilibrium properties, Ref.~\cite{Zohar2013} focused on an ultracold-atom realization, but without outlining an experimentally feasible protocol for gauge protection, which is crucial in analog quantum simulators where gauge symmetry-breaking errors are unavoidable \cite{Halimeh2020a,Halimeh_BriefReview}.

When it comes to quantum simulation experiments of $2+1$D QED or formulations thereof, there is an analog Rydberg realization where the electric and gauge fields are represented as qubits \cite{gonzalezcuadra2024observationstringbreaking2}. The plaquette term in this experiment is negligible, which is a common problem in analog quantum simulators, where the plaquette term emerges as a subleading term in degenerate perturbation theory. There is also a digital realization of a single plaquette using a qudit quantum processor, where the electric and gauge fields are natively represented as qudits \cite{Meth2025simulating}. This experiment focused mainly on a variational circuit to study ground-state properties, complemented by a small-scale demonstration of particle production dynamics.

Qudits have recently emerged as a powerful paradigm in quantum computing \cite{Ringbauer_2022}. Whereas qubits are two-level systems, qudits encode information in $d\geq2$ levels. This leads to great advantage in many areas of physics where local degrees of freedom are beyond binary. Qudits also have the potential to enable more efficient quantum gates, encoding them more compactly than qubits can. Given that LGTs can host long-range terms and gauge fields with large local Hilbert spaces, qudits become a natural choice in their quantum simulation, as evidenced by many recent proposals \cite{ciavarella2022conceptualaspectsoperatordesign,Popov2024variational,Calajo2024digital,kürkçüoglu2024quditgatedecompositiondependence,ballini2025symmetryverificationnoisyquantum,gaz2025quantumsimulationnonabelianlattice,jiang2025nonabeliandynamicscubeimproving}. As we show in this work, when it comes to $2+1$D spin-$S$ $\mathrm{U}(1)$ QLMs with dynamical hardcore-bosonic matter, qudits are particularly well-suited especially after integrating out the matter degrees of freedom through Gauss's law, which leads to long-range terms.

In this paper, we contribute to the development of scalable and resource-efficient implementations for simulating $2+1$D QED on near-term quantum hardware. We achieve this by integrating out the matter fields, and constructing a gauge-only (matter-integrated-out) Hamiltonian that admits a natural implementation using qudits; see Fig.~\ref{fig:lattice_scheme}. In addition, we propose a general approach for constructing circuits that support higher-spin representations, broadening the applicability of our framework. We benchmark our circuits by investigating the real-time dynamics of gauge fields in a nonperturbative regime over different system sizes. Simulating under realistic noise modeling --- dephasing and depolarization --- we find that coherent gauge-invariant dynamics persist for multiple oscillations even without any noise mitigation, e.g., without post-selection. This highlights the robustness and near-term viability of our approach, especially when some error mitigation is included.

The remainder of the paper is organized as follows. In Section \ref{Sec:ModelandInt}, we introduce the target Hamiltonian and demonstrate how matter fields can be systematically integrated out for arbitrary spin $S$ in arbitrary dimensions. In Section \ref{Sec:QuditCirc}, we construct qudit-based quantum circuits for implementing the full Hamiltonian dynamics in $2+1$D spin-$1/2$ case. In Section \ref{Sec:NumericalResults}, we present the results of numerical simulations based on Trotterized time evolution, comparing both noisy and noiseless settings. In Section \ref{Sec:Higher_spin_circ}, we present a general approach for constructing coupling term circuits for higher spins, with an explicit construction for the spin-$1$ case that is compatible with current quantum hardware. We conclude and provide outlook in Sec.~\ref{Conclusions}.

\section{Model and integration}
\label{Sec:ModelandInt}
In this work, we adopt a QLM formulation \cite{Chandrasekharan1997,Wiese_review} of $2+1$D scalar QED, incorporating a particle-hole transformation \cite{Hauke2013}. The Hamiltonian describing this model is \cite{Hashizume2022,osborne2022largescale21dmathrmu1gauge}
\begin{align}\nonumber
    \hat{H}_{\mathrm{QLM}} =&\, \hat{H}_C + \hat{H}_M
    + \hat{H}_E + \hat{H}_\square\\\nonumber
    =& -\kappa \sum_{\mathbf{r}, \mathbf{\nu}} \left(\hat{\phi}_\mathbf{r}^{\dag}\hat{s}_{\mathbf{r},\mathbf{e_{\nu}}}^-\hat{\phi}_{\mathbf{r+e_{\nu}}}^{\dag} + \text{H.c.}\right) + m\sum_\mathbf{r}\hat{\phi}_\mathbf{r}^{\dag}\hat{\phi}_\mathbf{r}\\ \label{h_qlm}
    &+ \frac{g^2}{2}\sum_{\mathbf{r}, \mathbf{\nu}}(\hat{s}_\mathbf{r, e_{\nu}}^z)^2 + J \sum_{\mathbf{r}}\left(\hat{U}_{\square_{\mathbf{r}}}+\hat{U}_{\square_{\mathbf{r}}}^{\dagger}\right),
\end{align}
where $\mathbf{r} = (r_x, r_y)^\intercal$ is the vector specifying the position of a lattice site and $\mathbf{e}_{\nu}$ is a unit vector along the direction of $\nu = (x, y)$. The matter fields, with mass $m$, are governed by the hardcore-bosonic ladder operators, $\hat{\phi}_\mathbf{r}^{\dagger}$ and $\hat{\phi}_\mathbf{r}$. Meanwhile, the gauge and electric fields, acting on the links between adjacent lattice sites, are represented by spin-$S$ operators \cite{Chandrasekharan1997,Wiese_review}, $\hat{s}_{\mathbf{r}, e_{\mathbf{\nu}}}^{\pm}$ and $\hat{s}_{\mathbf{r}, e_{\mathbf{\nu}}}^z$, respectively, simplifying the model while preserving the fundamental gauge symmetry.

The Hamiltonian consists of four terms. The minimal coupling
term $\hat{H}_C$, with a coupling strength $\kappa$, mediates matter exchange between neighboring sites while flipping the electric field in accordance with Gauss's law. The mass term $\hat{H}_M$ accounts for the mass energy of the matter fields. In addition, there are two pure gauge terms: $\hat{H}_E$, which captures the electric field energy, and the plaquette interaction $\hat{H}_\square$, involving the plaquette operator $\hat{U}_{\square_{\mathbf{r}}} = \hat{s}_{\mathbf{r}, \mathbf{e}_x}^{+} \hat{s}_{\mathbf{r}+\mathbf{e}_x, \mathbf{e}_y}^{-} \hat{s}_{\mathbf{r}+\mathbf{e}_y, \mathbf{e}_x}^{+} \hat{s}_{\mathbf{r}, \mathbf{e}_y}^{-}$, which governs the magnetic interactions for the gauge fields with an interaction strength $J$. 

The generator of the $ \mathrm{U}(1) $ gauge symmetry for the Hamiltonian $ \hat{H}_{\mathrm{QLM}} $ \eqref{h_qlm} is expressed as
\begin{align}\label{gauss}
    \hat{G}_\mathbf{r} = (-1)^{r_x + r_y}\left[\hat{\phi}_\mathbf{r}^{\dag}\hat{\phi}_\mathbf{r} + \sum_{\nu} \left(\hat{s}_\mathbf{r, e_{\nu}}^z + \hat{s}_\mathbf{r-e_{\nu}, e_{\nu}}^z\right)\right],
\end{align}
which corresponds to the discretized version of Gauss’s law. The commutation relations $[\hat{H}_{\mathrm{QLM}}, \hat{G}_\mathbf{r}] = 0,\, \forall \mathbf{r}$, encode the gauge invariance of the Hamiltonian, ensuring that the system preserves $\mathrm{U}(1)$ gauge symmetry. Here, we work in the physical sector of Gauss's law, i.e., we consider only those gauge-invariant states $\ket{\psi}$ that satisfy $\hat{G}_\mathbf{r}\ket{\psi} = 0$.

Since we are working with hardcore bosons, each matter site can accommodate at most one boson, or none, which imposes the constraint 
\begin{equation}
    \displaystyle\sum_{i \in \mathcal{N}(\mathbf{r})} s^z_i \in \{0, -1\},
\end{equation}
where $\mathcal{N}(\mathbf{r})$ denotes the set of links neighboring site $\mathbf{r}$ and $s^z_i$ is the eigenvalue of $\hat{s}^z_i$, 
on link $i$. Now, consider a shared link between two matter sites $\mathbf{r}$ and $\mathbf{r+e_{\nu}}$. The action of the term $\hat{H}_C$ in the Hamiltonian \eqref{h_qlm} corresponds to the correlated creation or annihilation of particles at the two matter sites, accompanied by the transition $\ket{s^z_{\mathbf{r, e_{\nu}}} = -m} \leftrightarrow\ket{s^z_{\mathbf{r, e_{\nu}}} = -m-1}$ on the shared electric field. Together with Gauss's law, this necessitates that the remaining neighboring electric fields for each matter site must collectively satisfy 
\begin{equation}
    \Bigg\vert \sum\limits_{\substack{i \in \mathcal{N}(\mathbf{r}) \\ \setminus \{ (\mathbf{r}, \mathbf{e}_\nu) \}}} s_i^z = m\Bigg\rangle  \,, \,\Bigg\vert\sum\limits_{\substack{i \in \mathcal{N} (\mathbf{r + e_{\nu}}) \\ \setminus \{ (\mathbf{r - e_{\nu}}, \mathbf{e}_\nu) \}}} s_i^z = m\Bigg\rangle,
\end{equation}
in order to survive the action of $\hat{H}_C$.

\begin{figure}[h]
    \hspace*{-0.02\textwidth}
\includegraphics[width=0.53\textwidth]{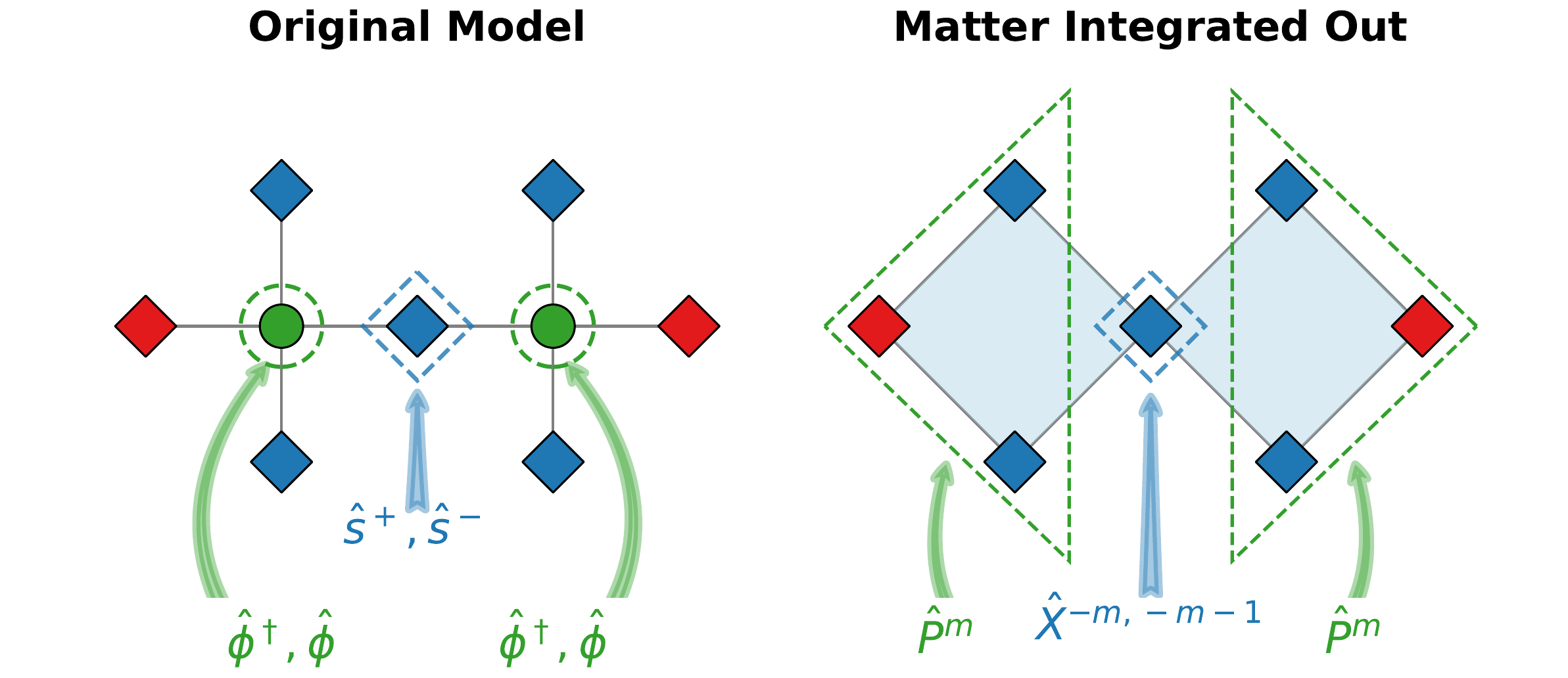}
\caption{Depiction of two neighboring vertices on the square lattice where the $2+1$D $\mathrm{U}(1)$ QLM is defined in the original model (left) and after integrating out the matter fields through Gauss's law (right). In the latter, the projector $\hat{P}^m_{\mathbf{r}}$ acts on the subsystem within the green triangles, while $\hat{X}^{-m,-m-1}_{\mathbf{r}, e_{\nu}}$ acts on the central electric field within the blue square.}\label{fig_MIO}
\end{figure}

We now formulate this in the picture where the matter fields are integrated out (Fig.~\ref{fig:lattice_scheme}). In this matter-integrated-out (MIO) picture, the links are denoted by their coordinates ($\mathbf{r}, \mathbf{e_{\nu}}$) as earlier. For each link ($\mathbf{r}, \mathbf{e_{\nu}}$), we define projectors that act on the electric fields surrounding the link. These projectors depend explicitly on the direction $\mathbf{\nu}$:

\begin{itemize}
    \item For $\mathbf{\nu} = \mathbf{x}$: The left projector $\hat{P}^{m}_{\mathbf{r}}$ acts on the set of links $L = \{(\mathbf{r}, \mathbf{e_{y}}), (\mathbf{r - e_{x}}, \mathbf{e_{x}}), (\mathbf{r-e_{y}}, \mathbf{e_{y}})\}$ while the right projector $\hat{P}^{m}_{\mathbf{r + e_{\nu}}}$ acts on the set of links $R = \{(\mathbf{r + e_{x}}, \mathbf{e_{x}}), (\mathbf{r + e_{x}}, \mathbf{e_{y}}), (\mathbf{r + e_{x} - e_{y}}, \mathbf{e_{y}})\}$.
    \item For $\mathbf{\nu} = \mathbf{y}$: The left projector $\hat{P}^{m}_{\mathbf{r}}$ acts on the set of links $L = \{(\mathbf{r}, \mathbf{e_{x}}), (\mathbf{r - e_{x}}, \mathbf{e_{x}}), (\mathbf{r-e_{y}}, \mathbf{e_{y}})\}$ and the right projector $\hat{P}^{m}_{\mathbf{r + e_{\nu}}}$ acts on the links $R = \{(\mathbf{r + e_{y}}, \mathbf{e_{y}}), (\mathbf{r + e_{y} - e_{x}}, \mathbf{e_{x}}), (\mathbf{r + e_{y})}, \mathbf{e_{x}}\}$.
\end{itemize}
These projectors project onto the subspace where the sum of the $s^z$ values over the specified surrounding links is $m$, and are defined explicitly as
\begin{equation}
\hat{P}^{m}_{\mathbf{r}} = \ketbra{\sum_{a \in L} s_a^z = m}_{\mathbf{r}},
\end{equation}
where the set $L$ is as defined earlier. The projector $\hat{P}^{m}_{\mathbf{r + e_{\nu}}}$ is defined analogously over the set $R$. The coupling term $\hat{H}_C$ in this formulation can be interpreted as the flipping Pauli operator $\hat{X}^{-m, -m-1}_{\mathbf{r}, \mathbf{e_{\nu}}}$ acting on the link ($\mathbf{r}, \mathbf{e_{\nu}}$) in the $\{\ket{-m}, \ket{-m-1}\}$ subspace, conditioned on the satisfaction of both projectors $\hat{P}^{m}_{\mathbf{r}}$ and $\hat{P}^{m}_{\mathbf{r + e_{\nu}}}$ on its neighboring links; see Fig.~\ref{fig_MIO}. For convenience, we define the rescaled operator $\hat{\sigma}^{x;-m, -m-1}_{\mathbf{r}, \mathbf{e_{\nu}}} = \sqrt{S(S+1) - m(m+1)} \, \hat{X}^{-m, -m-1}_{\mathbf{r}, \mathbf{e_{\nu}}}$, making our notation compact.
Thus, by integrating out the matter fields, the Hamiltonian \eqref{h_qlm} can be rewritten as
\begin{equation}\label{h_qudit_qlm}
\begin{aligned}
\hat{H}_{\mathrm{MIO}}^S ={}& -\kappa 
\sum_{m=-S}^{S-1} \sum_{\mathbf{r}, \mathbf{\nu}} \hat{P}^{m}_{\mathbf{r}} \, \hat{\sigma}^{x;-m, -m-1}_{\mathbf{r}, \mathbf{e_{\nu}}} \, \hat{P}^{m}_{\mathbf{r + e_{\nu}}} \\
& - 2m \sum_{\mathbf{r}, \mathbf{\nu}} \hat{s}^z_{\mathbf{r}, \mathbf{e_{\nu}}}  + \frac{g^2}{2} \sum_{\mathbf{r}, \mathbf{\nu}} (\hat{s}^z_{\mathbf{r}, \mathbf{e_{\nu}}})^2 \\
& + J \sum_{\mathbf{r}}\left(\hat{U}_{\square_{\mathbf{r}}}+\hat{U}_{\square_{\mathbf{r}}}^{\dagger}\right).
\end{aligned}
\end{equation}

It is important to note that this formulation generalizes to any number of spatial dimensions, where for a $d+1$D QLM, the projectors are defined over $(2d - 1)$ links, excluding the
shared link ($\mathbf{r, e_{\nu}}$).

In the following subsections, we consider the spin-$1/2$ QLM, for which the allowed configurations at each matter site and its four neighboring electric fields are shown in Fig.~\ref{fig:lattice_scheme}, along with their equivalent representation in the MIO formulation. For spin-$1/2$ QLM, the coupling term involves only $m = -1/2$, so we drop $m$ from operator superscripts. The matter-integrated-out Hamiltonian becomes
\begin{equation}\label{h_mio_s=1/2}
\begin{aligned}
\hat{H}_{\mathrm{MIO}}=& -\kappa 
\sum_{\mathbf{r}, \mathbf{\nu}} \hat{P}_{\mathbf{r}} \, \hat{\sigma}^{x}_{\mathbf{r}, \mathbf{e_{\nu}}} \, \hat{P}_{\mathbf{r + e_{\nu}}}- 2m \sum_{\mathbf{r}, \mathbf{\nu}} \hat{s}^z_{\mathbf{r}, \mathbf{e_{\nu}}}\\
& + \hat{H}_E + \hat{H}_\square,
\end{aligned}
\end{equation}
with projectors
\begin{equation}\label{projector_condition_s=1/2}
\hat{P}_{\mathbf{r}} = \ketbra{\sum_{a \in L} {s}_a^{z} = - \frac{1}{2}}_{\mathbf{r}}.
\end{equation}
In this case, $\hat{H}_E$ contributes only a constant energy and is therefore omitted.

\section{Qudit-based quantum circuits}
\label{Sec:QuditCirc}
For higher-spin systems ($S > 1/2$) in the formulation \eqref{h_qudit_qlm}, the intrinsic multilevel structure of qudits \cite{Ringbauer_2022} provides a natural framework for implementation, where different spin states can be encoded using distinct energy levels. The native gate set supports the required single-qudit operations and entangling interactions between different qudit levels. Moreover, qudits also offer an efficient approach to simulating the spin-$1/2$ QLM by eliminating the need for ancillary qubits and reducing the number of required entangling gates. In the following, we propose such an implementation for the spin-$1/2$ case by leveraging these unique properties of the trapped-ion qudit architecture.

To simulate time evolution under $\hat{H}_{\mathrm{MIO}}$ \eqref{h_mio_s=1/2}, we use the first-order Suzuki-Trotter decomposition,
\begin{equation}\label{u_trotter}
\hat{U}_{\rm ST}(\theta)
\;=\;
\Biggl(
\prod_{\mathbf{r}, \mathbf{\nu}}
\underbrace{e^{\imath 2m\hat{s}_{\mathbf{r, e_{\nu}}}^z d\theta}}_{\displaystyle \hat{U}_{M_{\mathbf{r, e_{\nu}}}}^{\rm MIO} (d\theta)}\,
\underbrace{e^{\imath \kappa\hat{P}_{\mathbf{r}}\, \hat{\sigma}^{x}_{\mathbf{r, e_{\nu}}} \hat{P}_{\mathbf{r+ e_{\nu}}} d\theta}}_{\displaystyle \hat{U}_{C_{\mathbf{r, e_{\nu}}}}^{\rm MIO}(d\theta)}\,
\underbrace{e^{-\imath \hat{H}_{\square_{\mathbf{r}}} d\theta}}_{\displaystyle \hat{U}_{\square_{\mathbf{r}}}^{\rm MIO}(d\theta)}
\Biggr)^{N}
\end{equation}
where $d\theta = \theta/N$ represents the size of a single Suzuki-Trotter step. In our simulations, we encode the spin states $\ket{s^z = -\frac{1}{2}}$ and $\ket{s^z = \frac{1}{2}}$ as $\ket{0}$ and $\ket{1}$ respectively. The first operator, $\hat{U}_{M_{\mathbf{r, e_{\nu}}}}^{\rm MIO}$ in $\hat{U}_{\mathrm{ST}}$ \eqref{u_trotter} can be easily implemented using single-qudit $R_z$ gates,
\begin{equation}\label{r_z}
\begin{aligned}
R_z^{ab}(\phi) = e^{-\imath \phi}\ketbra{a} + e^{\imath \phi}\ketbra{b},
\end{aligned}
\end{equation}
within the $\{\ket{0}, \ket{1}\}$ subspace of each qudit. The $R_z$ gates can be decomposed into two virtual $R_z$ gates. In the case of qudits, these generalize to the form
\begin{equation}
\text{VR}_z^{a}(\theta) = e^{-\imath\phi \ketbra{a}}. \label{virz}
\end{equation}
These gates are termed virtual because they are implemented through classical frame updates rather than physical pulses, and thus do not introduce any physical noise. Consequently, we can safely omit noise modeling for them in our simulations.
The quantum circuits for implementing the remaining two terms are presented in the following subsections.

\subsection{Realization of coupling term}
To simulate the evolution operator for the coupling term given in \eqref{u_trotter}, we use the qubit identities $e^{-\imath  \theta \hat{\sigma}^{x}_{\mathbf{r, e_{\nu}}}} = H_{\mathbf{r, e_{\nu}}} \, e^{-\imath  2\theta \hat{s}^{z}_{\mathbf{r, e_{\nu}}}}\, H_{\mathbf{r, e_{\nu}}}$, where $H_{\mathbf{r, e_{\nu}}}$ denotes the Hadamard operator, and
$(\hat{P}_{\mathbf{r}} \hat{\sigma}^x_{\mathbf{r, e_{\nu}}} \hat{P}_{\mathbf{r+e_{\nu}}})^2 = \hat{I}_{\hat{P}_{\mathbf{r}}} \hat{I}_{\mathbf{r, e_{\nu}}} \hat{I}_{\hat{P}_{\mathbf{r+e_{\nu}}}} = \hat{I}'
$, where each $\hat{I}$ represents the identity operator restricted to the relevant subspace defined by either the projectors, or $\hat{\sigma}^x_{\mathbf{r, e_{\nu}}}$ acting on the relevant qudits. Consequently, we have $\hat{{I}}'$ denoting the identity operator on the combined subspace.
These identities can be used to facilitate an efficient circuit decomposition:
\begin{equation}\label{unitary_coupling}
\begin{aligned}
\hat{U}_{C_{\mathbf{r, e_{\nu}}}}^{\rm MIO}(\theta) = & \sum_{n=0}^\infty\frac{\left(-\imath \theta \hat{P}_{\mathbf{r}} \hat{\sigma}^x_{\mathbf{r, e_{\nu}}} \hat{P}_{\mathbf{r+e_{\nu}}}\right)^n}{n!}\\
= &\, \mathbb{1} - \hat{I}' + \cos\theta \hat{I}' - \imath \sin\theta \, \hat{P}_{\mathbf{r}} \hat{\sigma}^x_{\mathbf{r, e_{\nu}}} \hat{P}_{\mathbf{r+e_{\nu}}}\\
= & \, \mathbb{1} - \hat{{I}}' + \hat{P}_{\mathbf{r}}\, (\cos\theta \,  \hat{{I}}_{\mathbf{r, e_{\nu}}}  - \imath \sin\theta \, \hat{\sigma}^x_{\mathbf{r, e_{\nu}}})\, \hat{P}_{\mathbf{r+e_{\nu}}}\\
= & \, \mathbb{1} - \hat{\mathrm{I}}' + \hat{P}_{\mathbf{r}}\, e^{-\imath \theta \hat{\sigma}^x_{\mathbf{r, e_{\nu}}}}\, \hat{P}_{\mathbf{r+e_{\nu}}}\\
= & \, \mathbb{1} - \hat{\mathrm{I}}' + \hat{P}_{\mathbf{r}}\, H_{\mathbf{r, e_{\nu}}} \, e^{-\imath 2\theta \hat{s}^z_{\mathbf{r, e_{\nu}}}}\, H_{\mathbf{r, e_{\nu}}} \, \hat{P}_{\mathbf{r+e_{\nu}}}.
\end{aligned}
\end{equation}
The evolution operator \eqref{unitary_coupling} acts nontrivially only on the states that survive the action of the $\hat{P}_{\mathbf{r}} \hat{\sigma}^x_{\mathbf{r, e_{\nu}}} \hat{P}_{\mathbf{r+e_{\nu}}}$ operator, where $\hat{P}_{\mathbf{r}} = \ketbra{001}+ \ketbra{010} + \ketbra{100}$ in qudit formulation. In our circuit, the neighboring qudits involved in the operators $\hat{P}_{\mathbf{r}}$ and $\hat{P}_{\mathbf{r+e_{\nu}}}$ are labeled as $\ket{P_L^k}$ and $\ket{P_R^k}$, respectively, with $k = 1, 2, 3$.

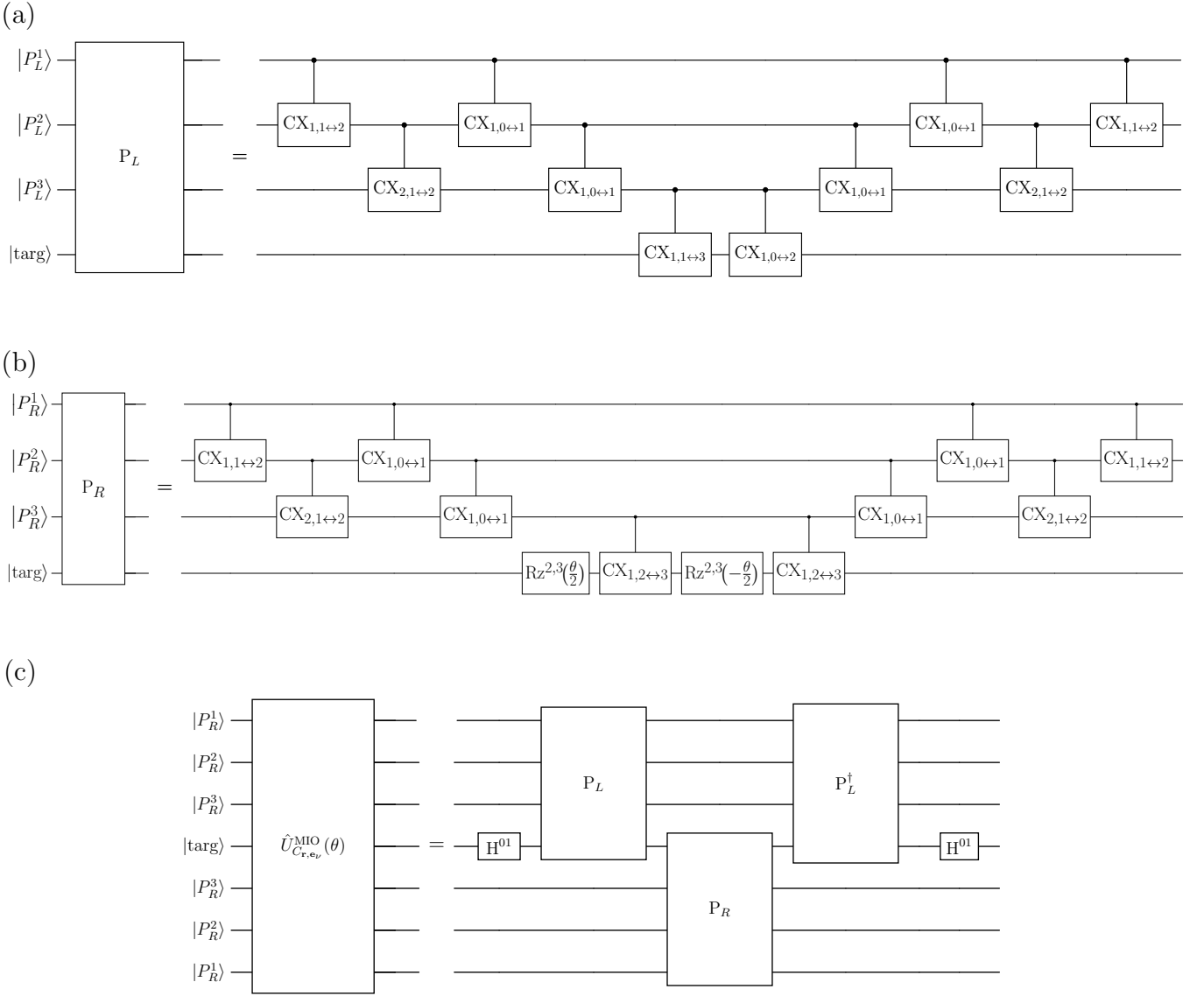
\begin{figure*}[t]
\captionsetup[subfloat]{position=top,justification=raggedright,singlelinecheck=false, font=large}
  \centering
  \subfloat[]{%
    \resizebox{\textwidth}{!}{%
\begin{quantikz}[row sep={1.8cm,between origins}, font=\Large]
  \lstick{$\ket{P_L^1}$} & \gate[4][3cm][1cm]{\mathrm{P}_{L}} & \qw & \midstick[4,brackets=none]{\textbf{\Large =}} & \ctrl{1} & & \ctrl{1} & & & & & \ctrl{1} & & \ctrl{1} &\\
  \lstick{$\ket{P_L^2}$} &                       & \qw &  & \gate[1][2cm][1.2cm]{\text{CX}_{1,1\leftrightarrow2}} 
                          & \ctrl{1} & \gate[1][2cm][1.2cm]{\text{CX}_{1,0\leftrightarrow1}} 
                          & \ctrl{1} & & & \ctrl{1} 
                          & \gate[1][2cm][1.2cm]{\text{CX}_{1,0\leftrightarrow1}} 
                          & \ctrl{1} 
                          & \gate[1][2cm][1.2cm]{\text{CX}_{1,1\leftrightarrow2}} 
                          &\\
  \lstick{$\ket{P_L^3}$} &                       & \qw &  & & \gate[1][2cm][1.2cm]{\text{CX}_{2,1\leftrightarrow2}} 
                         & & \gate[1][2cm][1.2cm]{\text{CX}_{1,0\leftrightarrow1}} 
                         & \ctrl{1} & \ctrl{1} 
                         & \gate[1][2cm][1.2cm]{\text{CX}_{1,0\leftrightarrow1}} 
                         & & \gate[1][2cm][1.2cm]{\text{CX}_{2,1\leftrightarrow2}} 
                         & &\\
  \lstick{$\ket{\text{targ}}$} &                   & \qw &  & & & & & \gate[1][2cm][1.2cm]{\text{CX}_{1,1\leftrightarrow3}} 
                          & \gate[1][2cm][1.2cm]{\text{CX}_{1,0\leftrightarrow2}} 
                          & & & & &\\
\end{quantikz}
    }%
  }
  \hfill
  \subfloat[]{%
    \resizebox{\textwidth}{!}{%
\begin{quantikz}[row sep={2.7cm,between origins}, font=\Huge]
  \lstick{$\ket{P_R^1}$} & \gate[4][3cm][1cm]{\mathrm{P}_{R}} & \qw & \midstick[4,brackets=none]{\textbf{\Huge =}} & \ctrl{1} & & \ctrl{1} & & & & & & & \ctrl{1} & & \ctrl{1} &\\
  \lstick{$\ket{P_R^2}$} &                       & \qw &  & \gate[1][2cm][2cm]{\text{CX}_{1,1\leftrightarrow2}} 
                          & \ctrl{1} 
                          & \gate[1][2cm][2cm]{\text{CX}_{1,0\leftrightarrow1}} 
                          & \ctrl{1} & & & & & \ctrl{1} 
                          & \gate[1][2cm][2cm]{\text{CX}_{1,0\leftrightarrow1}} 
                          & \ctrl{1} 
                          & \gate[1][2cm][2cm]{\text{CX}_{1,1\leftrightarrow2}} &\\
  \lstick{$\ket{P_R^3}$} &                       & \qw &  & & \gate[1][2cm][2cm]{\text{CX}_{2,1\leftrightarrow2}} 
                         & & \gate[1][2cm][2cm]{\text{CX}_{1,0\leftrightarrow1}} 
                         & & \ctrl{1} 
                         & & \ctrl{1} 
                         & \gate[1][2cm][2cm]{\text{CX}_{1,0\leftrightarrow1}} 
                         & & \gate[1][2cm][2cm]{\text{CX}_{2,1\leftrightarrow2}} 
                         & &\\
  \lstick{$\ket{\text{targ}}$} &                   & \qw &  & & & & & \gate[1][2cm][2cm]{\mathrm{Rz}^{2,3}\!\bigl(\tfrac\theta2\bigr)}  
                          & \gate[1][2cm][2cm]{\text{CX}_{1,2\leftrightarrow3}} 
                          & \gate[1][2cm][2cm]{\mathrm{Rz}^{2,3}\!\bigl(-\tfrac\theta2\bigr)}  
                          & \gate[1][2cm][2cm]{\text{CX}_{1,2\leftrightarrow3}} 
                          & & & & &\\
\end{quantikz}
    }%
  }\hfill
  \captionsetup[subfloat]{oneside,margin={-2.6cm,0cm}}
  \subfloat[]{%
    \resizebox{0.7\textwidth}{!}{%
    \begin{quantikz}[row sep={1cm,between origins}, font=\large]
    \lstick{$\ket{P_R^1}$} & \gate[7][2.9cm][1cm]{\hat{U}_{C_{\mathbf{r,e_{\nu}}}}^{\rm MIO}(\theta)} & \qw & \midstick[7,brackets=none]{\Large$\mathbf{=}$} & & \gate[4][2.5cm]{\mathrm{P}_{L}} & & \gate[4][2.5cm]{\mathrm{P}^{\dag}_{L}} & & &\\
    \lstick{$\ket{P_R^2}$} & & \qw & & & & & & & &\\
    \lstick{$\ket{P_R^3}$} & & \qw & & & & & & & &\\
    \lstick{$\ket{\text{targ}}$} & & \qw & & \gate[1][1cm]{\text{H}^{01}} & & \gate[4][2.5cm]{\mathrm{P}_{R}} & & & \gate[1]{\text{H}^{01}} &\\
    \lstick{$\ket{P_R^3}$} & & \qw & & & & & & & &\\
    \lstick{$\ket{P_R^2}$} & & \qw & & & & & & & &\\
    \lstick{$\ket{P_R^1}$} & & \qw & & & & & & & &\\
\end{quantikz}
}
}

  \caption{(a) and (b) depict the subcircuits $\mathrm{P}_L$ and $\mathrm{P}_R$, which verify the projectors $\hat{P}_{\mathbf{r}}$ and $\hat{P}_{\mathbf{r + e_\nu}}$, respectively. These subcircuits enable the desired transitions described in the main text, or the application of a controlled–$Z$ rotation on the target qudit within the ${\ket{2}, \ket{3}}$ subspace, only when the projector condition~\eqref{projector_condition_s=1/2} is satisfied. Panel (c) shows the complete quantum circuit for $\hat{U}_{C_{\mathbf{r, e_\nu}}}^{\mathrm{MIO}}(\theta)$. The full circuit comprises 22 two-qudit gates and four single-qudit gates, with some gates in $\mathrm{P}_L$ and $\mathrm{P}_L^{\dagger}$ canceling out to the identity. $\mathrm{H}^{01}$ denotes the Hadamard gate acting in the ${\ket{0}, \ket{1}}$ subspace.}
  \label{fig:H_C Circuit}
\end{figure*}

\begin{table}[H]
    \centering
    \begin{tabular}{|c|c|}
        \hline
        \textbf{3-link state} & \textbf{3-qudit state} \\  
        \hline
        $\ket{000}$ & $\ket{000}$ \\  
        \hline
        $\ket{001}$ & $\ket{001}$ \\  
        \hline
        $\ket{010}$ & $\ket{011}$ \\  
        \hline
        $\ket{011}$ & $\ket{010}$ \\  
        \hline
        $\ket{100}$ & $\ket{111}$ \\  
        \hline
        $\ket{101}$ & $\ket{110}$ \\  
        \hline
        $\ket{110}$ & $\ket{120}$ \\  
        \hline
        $\ket{111}$ & $\ket{122}$ \\  
        \hline
    \end{tabular}
    \caption{Mapping of 3-link states for spin-$\frac{1}{2}$ model: Only the states $\ket{001}$, $\ket{010}$, and $\ket{100}$ are mapped to configurations where the last qudit is in the state $\ket{1}$.}
    \label{tab:1}
\end{table}

Since qudits provide a $ d $-dimensional Hilbert space, a system of three qudits spans a total of $ d^3 $ states. In our approach, we encode spin-$1/2$ systems into qudits with a maximum dimension of 4, embedding the relevant set of eight three-link states within a larger Hilbert space of dimension $ 4^3 $. This embedding allows for a structured mapping of three-link states onto three-qudit states, as presented in Table~\ref{tab:1}. Given the projector operator definition, we choose a mapping that ensures the last qudit remains in the $\ket{1}$ state for the relevant three-link states. For all other three-link states, this condition is not satisfied. This design enables an efficient circuit implementation, requiring only four CX gates to realize the necessary transformations.

The circuit begins with the $\mathrm{H}^{01}$ gate -- representing the Hadamard gate restricted to the $\{\ket{0}, \ket{1}\}$ subspace of a qudit -- applied to the target qudit, creating a superposition of the $\ket{0}$ and $\ket{1}$ states. Subsequently, the projector $\hat{P}_{\mathbf{r}}$ \eqref{projector_condition_s=1/2} is implemented using the dedicated subcircuit $\mathrm{P}_L$, shown in Fig.~\ref{fig:H_C Circuit}(a), ensuring that the state $\ket{P_L^3}$ transitions to $\ket{1}$ only if the state $\ket{P_L^1 P_L^2 P_L^3}$ satisfies the conditions imposed by $\hat{P}_{\mathbf{r}}$.

The qudit $\ket{P_L^3}$ then functions as the control qudit, enabling the state transitions $\ket{0} \rightarrow \ket{2}$ and $\ket{1} \rightarrow \ket{3}$ in the target qudit, provided that $\ket{P_L^3}$ is in the $\ket{1}$ state. If these conditions are not met, the target qudit remains unchanged. This subcircuit is realized entirely through controlled exchange (CX) gates \cite{Ringbauer_2022}, denoted as $\mathrm{CX}_{c, l_1 \leftrightarrow l_2}$ which apply an $X$ gate between the $\ket{l_1}$ and $\ket{l_2}$ states of the target qudit, conditioned on the control qudit being in state $\ket{c}$,
\begin{equation}\label{CX}
\begin{aligned}
\operatorname{CX}_{c, l_1 \leftrightarrow l_2}:\left\{
\begin{array}{l}
\ket{c, l_1} \leftrightarrow \ket{c, l_2} \\
\ket{j, k} \rightarrow \ket{j, k} \quad \text{for } j \neq c, \, k \neq l_1, l_2
\end{array}
\right. 
\end{aligned}
\end{equation}

Similarly, it is ensured through the subcircuit $\mathrm{P_R}$, shown in Fig.~\ref{fig:H_C Circuit}(b), that the state $\ket{P_R^3}$ transitions to $\ket{1}$ only if $\ket{P_R^1 P_R^2 P_R^3}$ satisfies the conditions imposed by $\hat{P}_{\mathbf{r+e_{\nu}}}$. When $\ket{P_R^3}$ is in the $\ket{1}$ state, a controlled-$Z$ rotation is applied to the target qudit within the $\{\ket{2}, \ket{3}\}$ subspace, using $\ket{P_R^3}$ as the control qudit. Subsequently, the subcircuit $\mathrm{P_L}$ is reversed, restoring all qudits to the $\{\ket{0}, \ket{1}\}$ subspace, followed by the application of another Hadamard gate. The Hadamard gates serve to facilitate the correct implementation of the controlled-$X$ rotation. This approach enables the control over $\theta$ to be applied using two single-qudit gates instead of multi-qudit controlled operations, significantly simplifying the implementation.

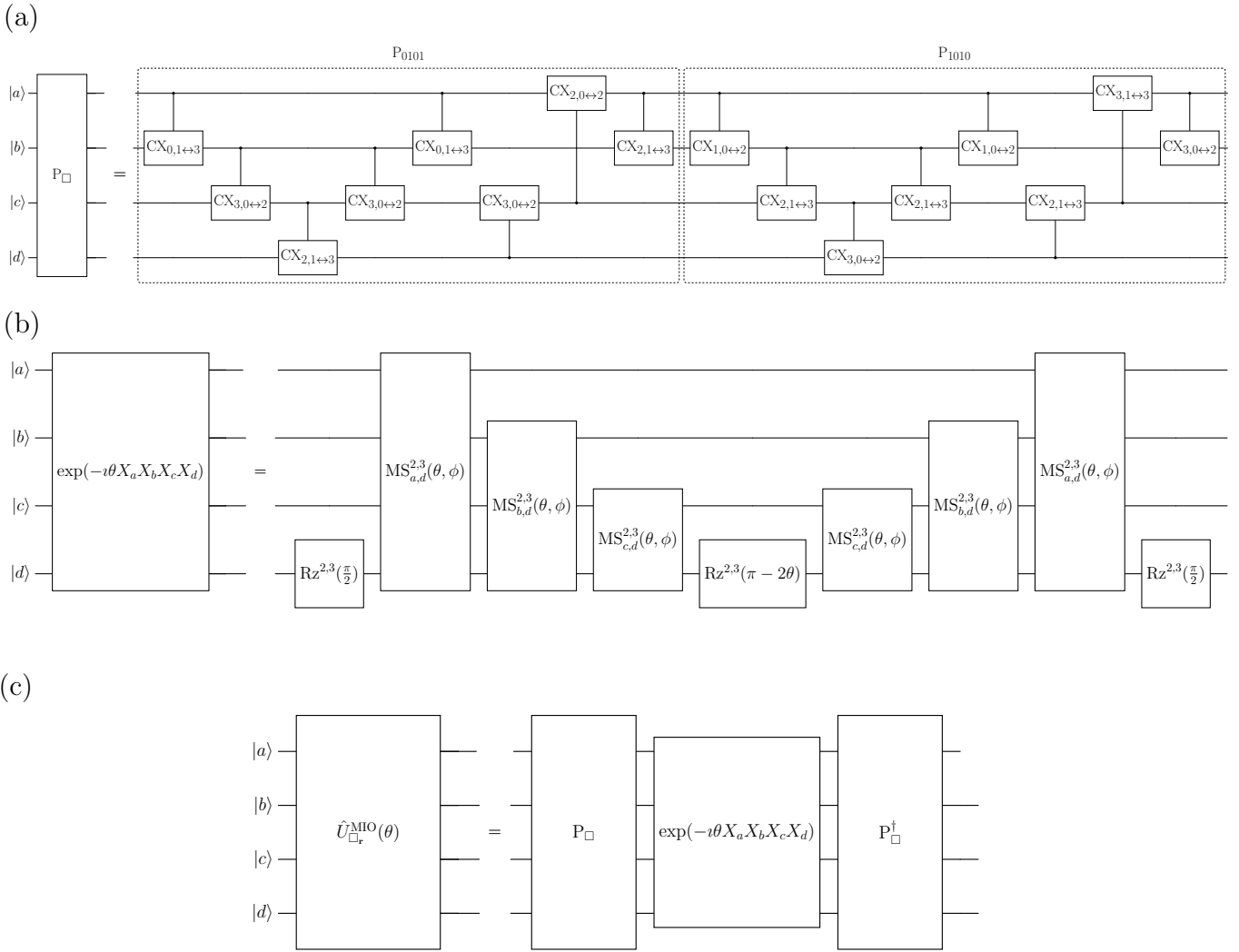
\begin{figure*}[t]
\captionsetup[subfloat]{position=top,justification=raggedright,singlelinecheck=false, font=large}
  \centering
  \subfloat[]{%
    \resizebox{\linewidth}{!}{%
\begin{quantikz}[row sep={3.2cm,between origins}, font=\Huge]
        \lstick{$\ket{a}$} & \gate[4][2.9cm][2.2cm]{\mathrm{P}_{\square}} & \qw & \midstick[4,brackets=none]{\textbf{\Huge =}} & \ctrl{1} \gategroup[wires=4, steps=8, style={dashed,rounded
corners, inner sep=7pt}, label style={label position=above, yshift=0.5cm}]{{$\mathrm{P}_{0101}$}} & & & & \ctrl{1}
                           & & \gate[1][2cm][2cm]{\text{CX}_{2,0\leftrightarrow2}} 
                           & \ctrl{1} & & \ctrl{1} \gategroup[wires=4, steps=8, style={dashed,rounded
corners, inner sep=7pt}, label style={label position=above, yshift=0.5cm}]{{$\mathrm{P}_{1010}$}} & & & & \ctrl{1} 
                           & & \gate[1][2cm][2cm]{\text{CX}_{3,1\leftrightarrow3}} 
                           & \ctrl{1} & \\
        \lstick{$\ket{b}$} & & \qw & & \gate[1][2cm][2cm]{\text{CX}_{0,1\leftrightarrow3}} 
                           & \ctrl{1} &  & \ctrl{1} 
                           & \gate[1][2cm][2cm]{\text{CX}_{0,1\leftrightarrow3}} 
                           & & & \gate[1][2cm][2cm]{\text{CX}_{2,1\leftrightarrow3}} 
                            & & \gate[1][2cm][2cm]{\text{CX}_{1,0\leftrightarrow2}} 
                           & \ctrl{1} &  & \ctrl{1} 
                           & \gate[1][2cm][2cm]{\text{CX}_{1,0\leftrightarrow2}} 
                           & & & \gate[1][2cm][2cm]{\text{CX}_{3,0\leftrightarrow2}} 
                           & \\
        \lstick{$\ket{c}$} & & \qw & & & \gate[1][2cm][2cm]{\text{CX}_{3,0\leftrightarrow2}} 
                           & \ctrl{1}  
                           & \gate[1][2cm][2cm]{\text{CX}_{3,0\leftrightarrow2}} 
                           & & \gate[1][2cm][2cm]{\text{CX}_{3,0\leftrightarrow2}} 
                           & \ctrl{-2} 
                          & & & & \gate[1][2cm][2cm]{\text{CX}_{2,1\leftrightarrow3}} 
                           & \ctrl{1}  
                           & \gate[1][2cm][2cm]{\text{CX}_{2,1\leftrightarrow3}} 
                           & & \gate[1][2cm][2cm]{\text{CX}_{2,1\leftrightarrow3}} 
                           & \ctrl{-2} 
                           & &\\
        \lstick{$\ket{d}$} & & \qw & & & & \gate[1][2cm][2cm]{\text{CX}_{2,1\leftrightarrow3}} 
                           & & & \ctrl{-1} 
                           & & & & & & \gate[1][2cm][2cm]{\text{CX}_{3,0\leftrightarrow2}} 
                           & & & \ctrl{-1} 
                           & & &
\end{quantikz}
    }%
  }\hfill
  \subfloat[]{%
    \resizebox{\linewidth}{!}{%
      \begin{quantikz}[row sep={2cm,between origins}, font=\Large]
    \lstick{$\ket{a}$} & \gate[4][2.5cm][1cm]{\mathrm{exp}(-\imath\theta X_a X_b X_c X_d)} & \qw & \midstick[4,brackets=none]{\textbf{\large =}} & & \gate[4][2cm][1cm]{\mathrm{MS}^{2,3}_{a,d}(\theta, \phi)} & & & & & &  \gate[4][2cm][1cm]{\mathrm{MS}^{2,3}_{a,d}(\theta, \phi)} & &\\
    \lstick{$\ket{b}$} & & \qw & & & &  \gate[3][2cm][1cm]{\mathrm{MS}^{2,3}_{b,d}(\theta, \phi)} & & & &  \gate[3][2cm][1cm]{\mathrm{MS}^{2,3}_{b,d}(\theta, \phi)} & & &  \\
    \lstick{$\ket{c}$} & & \qw & & & & &  \gate[2][2cm][1cm]{\mathrm{MS}^{2,3}_{c,d}(\theta, \phi)} & &  \gate[2][2cm][1cm]{\mathrm{MS}^{2,3}_{c,d}(\theta, \phi)} & & & & \\
    \lstick{$\ket{d}$} & & \qw & & \gate[1][2cm][2cm]{\text{Rz}^{2,3} (\frac{\pi}{2})} & & & & \gate[1][2cm][2cm]{\text{Rz}^{2,3} (\pi - 2\theta)} &  & & & \gate[1][2cm][2cm]{\text{Rz}^{2,3} (\frac{\pi}{2})} & \\
\end{quantikz}
    }%
  }\hfill
  \captionsetup[subfloat]{oneside,margin={-3.6cm,0cm}}
  \subfloat[]{%
    \resizebox{0.6\linewidth}{!}{%
    \begin{quantikz}[row sep={1.5cm,between origins}, font=\Large]
    \lstick{$\ket{a}$} & \gate[4][4cm][2cm]{\hat{U}_{\square_{\mathbf{r}}}^{\rm MIO} (\theta)} & \qw & \midstick[4,brackets=none]{\textbf{\large =}} & \gate[4][2.9cm][2cm]{\mathrm{P}_{\square}} & \gate[4]{\mathrm{exp}(-\imath\theta X_a X_b X_c X_d)} & \gate[4][2.9cm][2cm]{\mathrm{P}^{\dag}_{\square}} &\\
    \lstick{$\ket{b}$} & & \qw & & & & & &\\
    \lstick{$\ket{c}$} & & \qw & & & & & &\\
    \lstick{$\ket{d}$} & & \qw & & & & & &\\
\end{quantikz}
}
}
  \caption{%
    (a) shows the subcircuit $\mathrm{P}_{\square}$ used to verify the plaquette projector $\hat{P}_{\square_{\mathbf{r}}}$ and to implement the transitions $\ket{0} \rightarrow \ket{2}$ and $\ket{1} \rightarrow \ket{3}$ when the verification succeeds. (b) presents the subcircuit $\exp(-\imath \theta\,X_a X_b X_c X_d)$, which decomposes the plaquette interaction term $e^{-\imath  \theta\hat{\sigma}^x_{\mathbf{r,e_x}} \hat{\sigma}^x_{\mathbf{r+e_x,e_y}} \hat{\sigma}^x_{\mathbf{r+e_y,e_x}} \hat{\sigma}^x_{\mathbf{r,e_y}}}$ into MS and $R_z$ gates. (c) shows the complete quantum circuit implementing the plaquette interaction, consisting of 38 entangling gates.
  }
\label{fig:H_plaq}
\end{figure*}

\subsection{Simulating the magnetic interactions}
The structure of the plaquette term ensures that the dynamics involves only two states, $\ket{0101}$ and $\ket{1010}$. Therefore, we can write
\begin{equation}
\begin{aligned}
\hat{H}_{\square_{\mathbf{r}}}^{\mathrm{MIO}} & = \ketbra{1010}{0101} + \ketbra{0101}{1010}\\
& =  \hat{\sigma}^x_{\mathbf{r,e_x}} \hat{\sigma}^x_{\mathbf{r+e_x,e_y}} \hat{\sigma}^x_{\mathbf{r+e_y,e_x}} \hat{\sigma}^x_{\mathbf{r,e_y}}
\hat{P}_{\square_{\mathbf{r}}},
\end{aligned}
\end{equation}
where $\hat{P}_{\square_{\mathbf{r}}} = \ketbra{1010} + \ketbra{0101}$. Additionally, it is important to note that $(\hat{H}_{\square_{\mathbf{r}}})^2 = \hat{{I}}_{\square_{\mathbf{r}}}$, the identity operator on the subspace spanned by the states $\ket{0101}$ and $\ket{1010}$. The evolution operator can then be written as
\begin{equation}\label{circ_plaq}
\begin{aligned}
\hat{U}_{\square_{\mathbf{r}}}^{\rm MIO}(\theta)
= & \sum_{n=0}^\infty\frac{\big({-}\imath  \theta \hat{H}_{\square_{\mathbf{r}}}^{\mathrm{MIO}}\big)^n}{n!}\\
= & \, \mathbb{1} - \hat{{I}}_{\square_{\mathbf{r}}} + \cos\theta \hat{{I}}_{\square_{\mathbf{r}}} - \imath  \sin\theta \hat{H}_{\square_{\mathbf{r}}}\\
= & \mathbb{1} - \hat{{I}}_{\square_{\mathbf{r}}} +  e^{-\imath \theta\hat{\sigma}^x_{\mathbf{r,e_x}} \hat{\sigma}^x_{\mathbf{r+e_x,e_y}} \hat{\sigma}^x_{\mathbf{r+e_y,e_x}} \hat{\sigma}^x_{\mathbf{r,e_y}}} \hat{P}_{\square_{\mathbf{r}}}.
\end{aligned}
\end{equation}
The circuit $\mathrm{P_{\square}}$ (Fig.~\ref{fig:H_plaq}(a)), implementing $\hat{P}_{\square_{\mathbf{r}}}$, is decomposed into two subcircuits: $\mathrm{P_{1010}}$, which maps the state $\ket{1010}$ to $\ket{3232}$, and $\mathrm{P_{0101}}$, which transforms $\ket{0101}$ into $\ket{2323}$. Any other state remains unaffected by the circuit. The evolution operator  
$e^{-\imath \theta\hat{\sigma}^x_{\mathbf{r,e_x}} \hat{\sigma}^x_{\mathbf{r+e_x,e_y}} \hat{\sigma}^x_{\mathbf{r+e_y,e_x}} \hat{\sigma}^x_{\mathbf{r,e_y}}}
$
is implemented through a subcircuit (Fig.~\ref{fig:H_plaq}(b)) using $R_z^{2,3} (\phi)$ and Mølmer–Sørensen (MS) gates, defined as
\begin{equation}
\mathrm{MS}^{a, b}(\theta, \phi)=\exp \left(-\frac{\imath  \theta}{4}\left[\sigma_\phi^{a, b} \otimes \mathbb{1}+\mathbb{1} \otimes \sigma_\phi^{a, b}\right]^2\right).
\end{equation}
where $\sigma^{i,j}_{\phi} = (\cos  \phi \,\sigma_x^{i,j}\pm \sin \phi \, \sigma_y^{i,j})$ for Pauli matrices $\sigma_x, \sigma_y$ and $\{i, j\}$ denotes the addressed subspace.
\begin{figure*}[t]
\captionsetup[subfloat]{position=top,justification=raggedright,singlelinecheck=false, font=large}
\centering
\subfloat[]{
\hspace*{-2em}
  \includegraphics[width=0.5\linewidth]{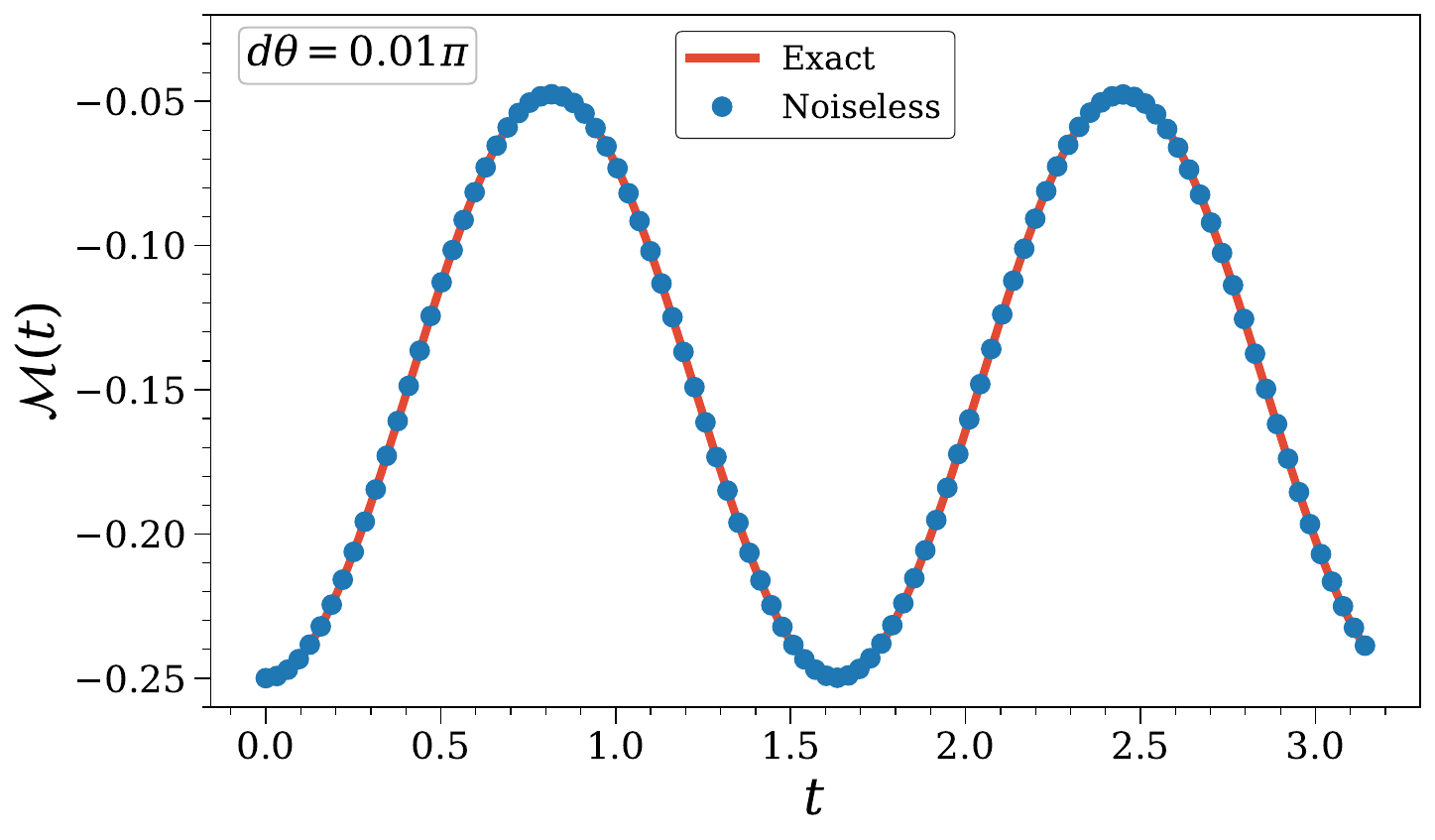}
}
\hfill
\subfloat[]{
  \includegraphics[width=0.5\linewidth]{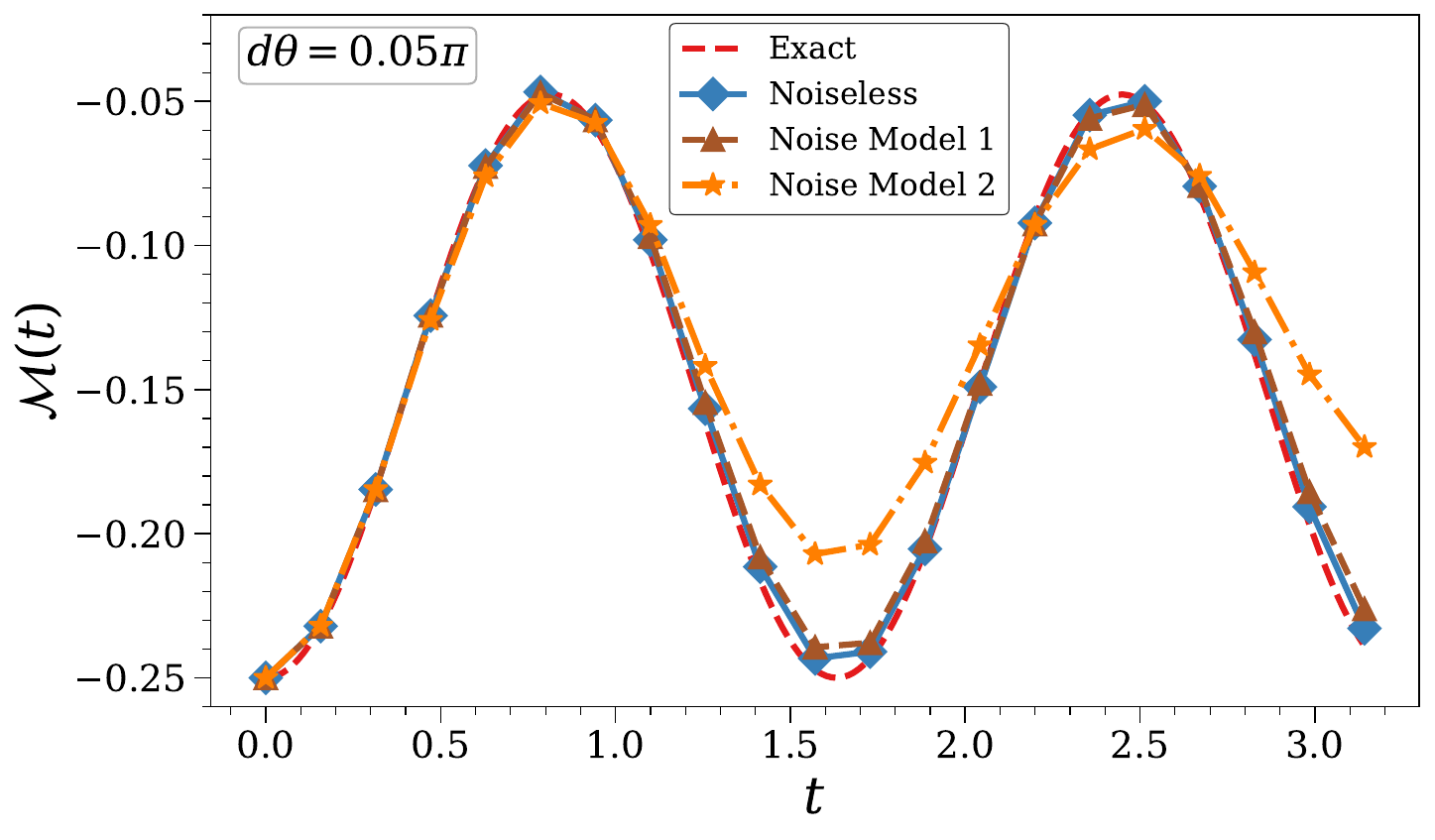}
}


\subfloat[]{
  \includegraphics[width=\linewidth]{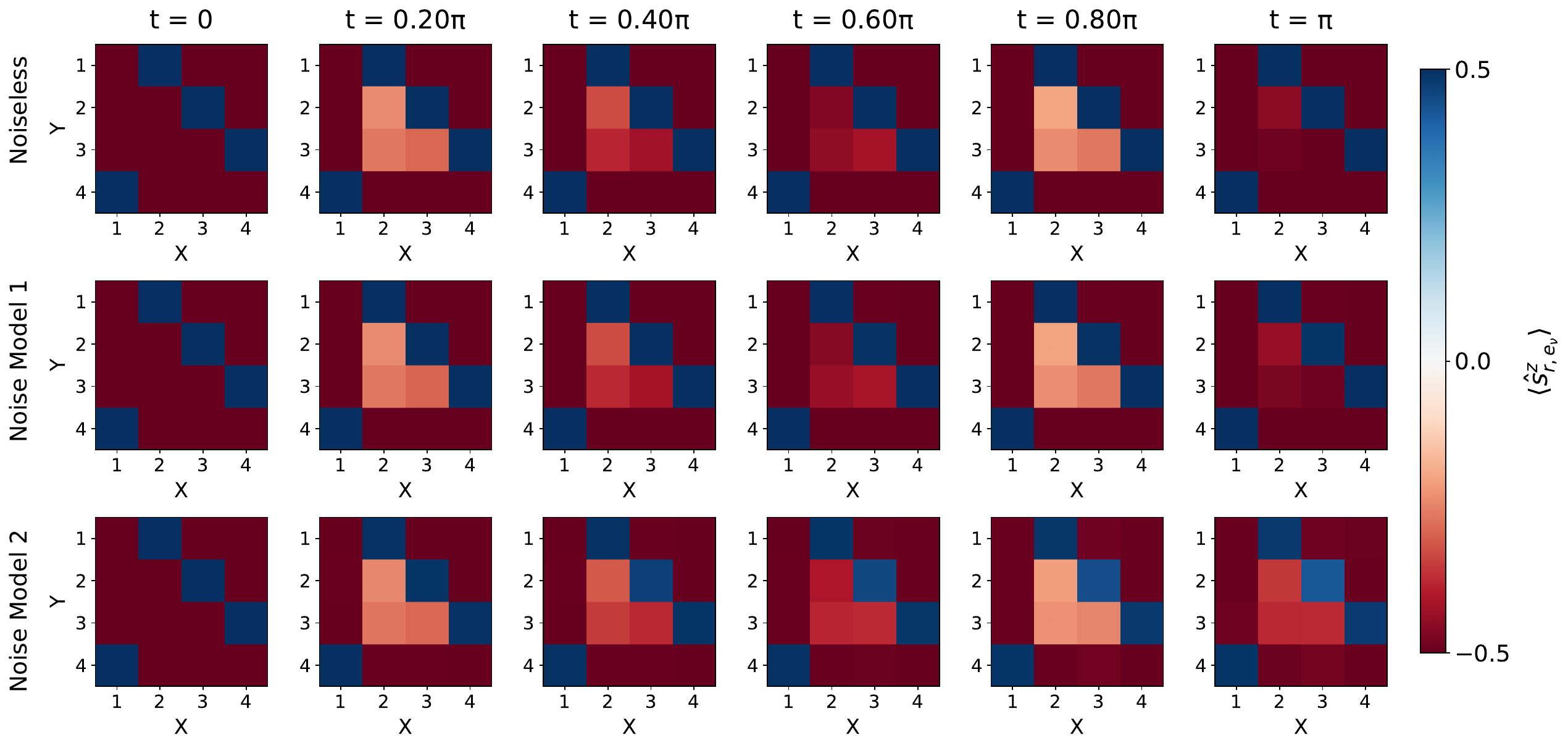}
}
\caption{Dynamics of $\mathcal{M}(t)$ over the four center spins in System I using parameters $m=0.42$, $\kappa = 1$ and $J=0$: (a) Comparison between noiseless Trotterized simulations and exact results using $d\theta = 0.01\pi$, (b) Comparison of noisy simulations against exact and noiseless results for $d\theta = 0.05\pi$, (c) Snapshots of the lattice at representative times, highlighting the local $\langle \hat{s}^z_{\mathbf{r, e_{\nu}}} \rangle$  dynamics under different noise conditions relative to the noiseless case.}
\label{fig:combined_magnetization_dynamics_snapshots}
\end{figure*}

The $\mathrm{MS}^{2, 3}(\theta, \phi)$ gate serves as a two-qudit entangling operation that acts exclusively within the $\{\ket{2}, \ket{3}\}$ subspace, ensuring that the subcircuit operates solely within the subspace spanned by $\{\ket{2323}, \ket{3232}\}$. Finally, each qudit state is mapped back to the $\{\ket{0}, \ket{1}\}$ basis to realize \eqref{circ_plaq} faithfully through the circuit shown in Fig.~\ref{fig:H_plaq}(c).

\begin{figure*}[t]
\captionsetup[subfloat]{position=top,justification=raggedright,singlelinecheck=false, font=large}
    \centering
    \subfloat[]{
    \hspace*{-2em}
        \includegraphics[width=0.5\textwidth]{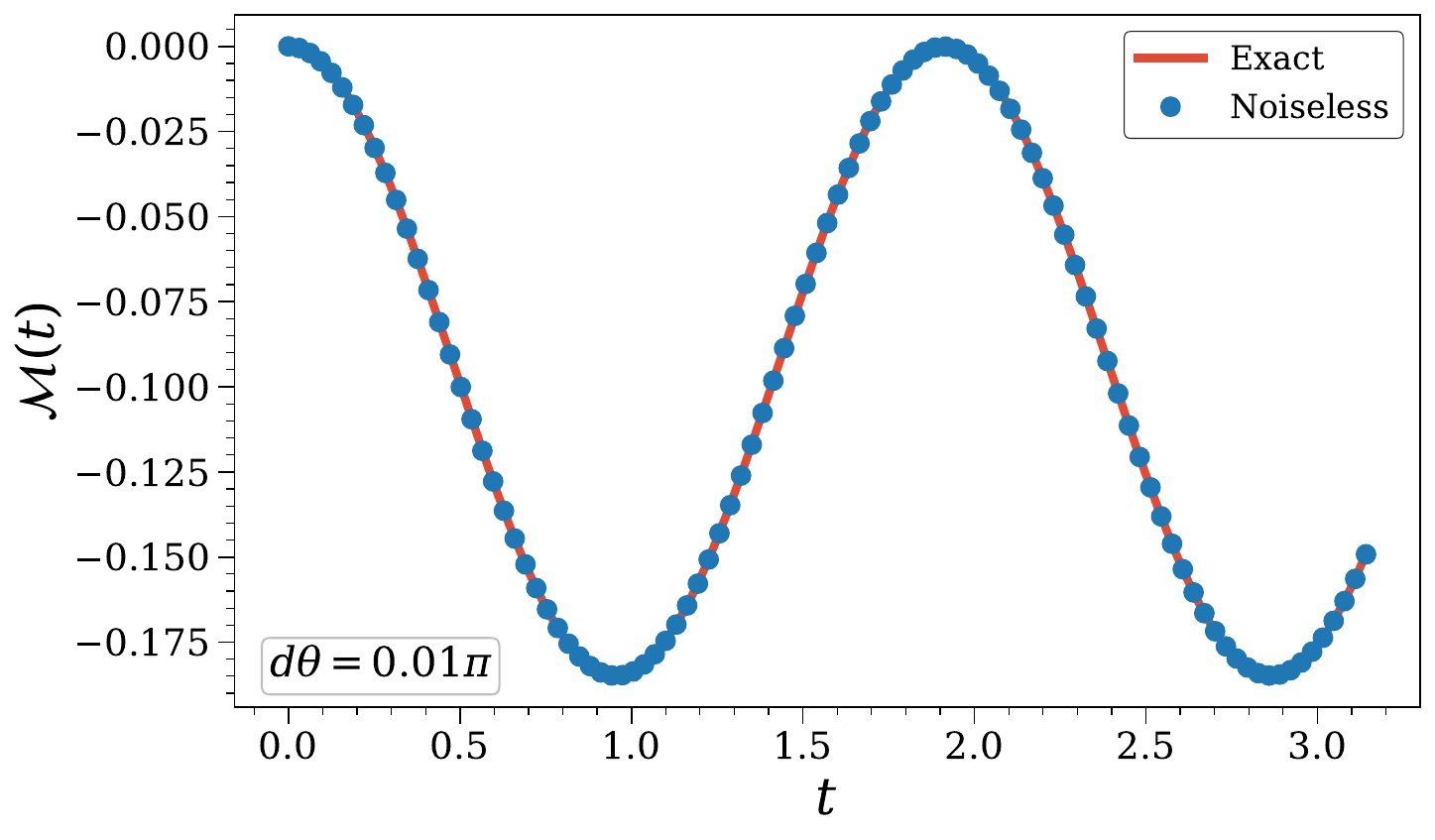}
    }
    \hfill
    \subfloat[]{
        \includegraphics[width=0.5\textwidth]{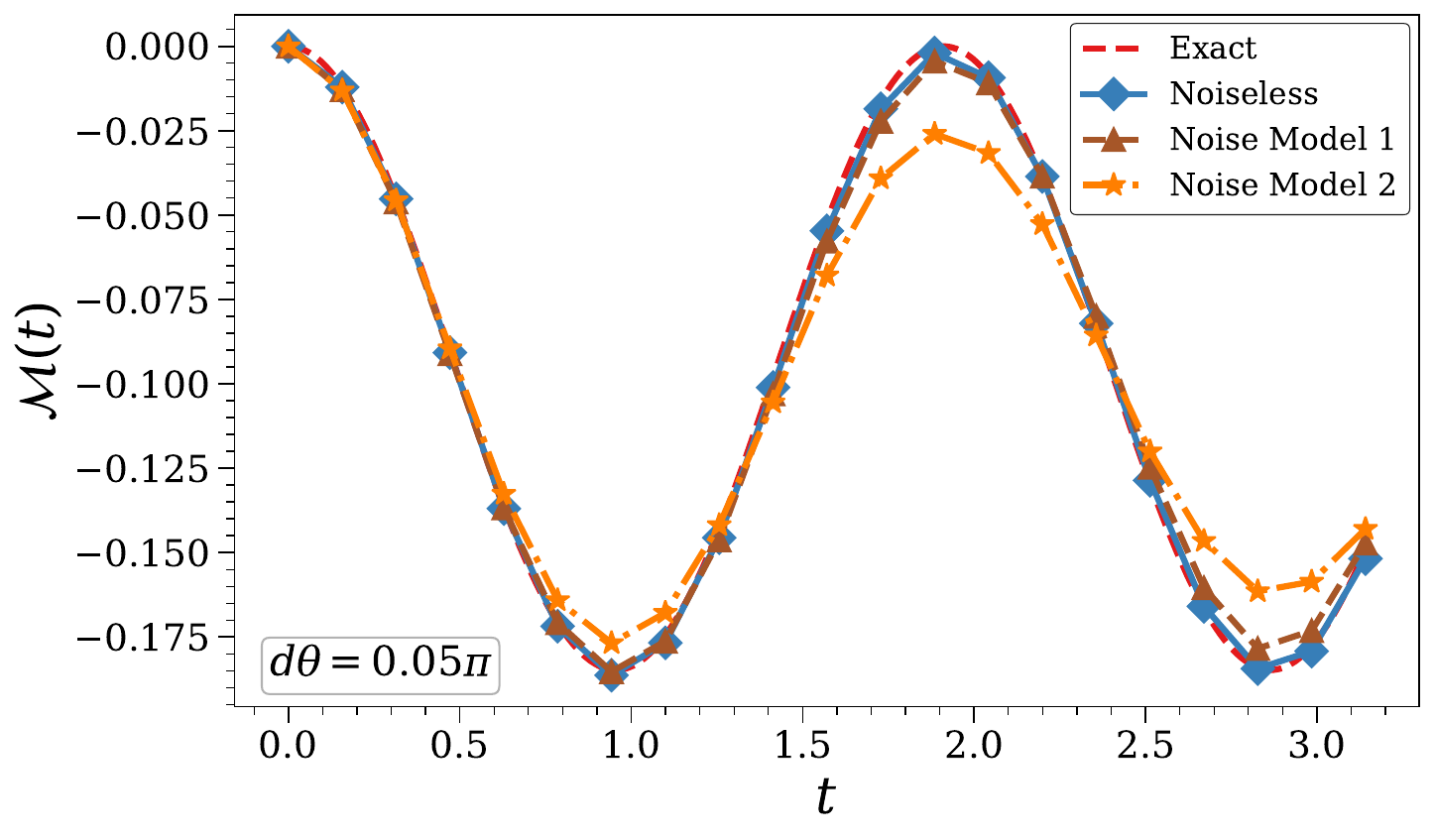}
    }


    \subfloat[]{
        \includegraphics[width=\textwidth]{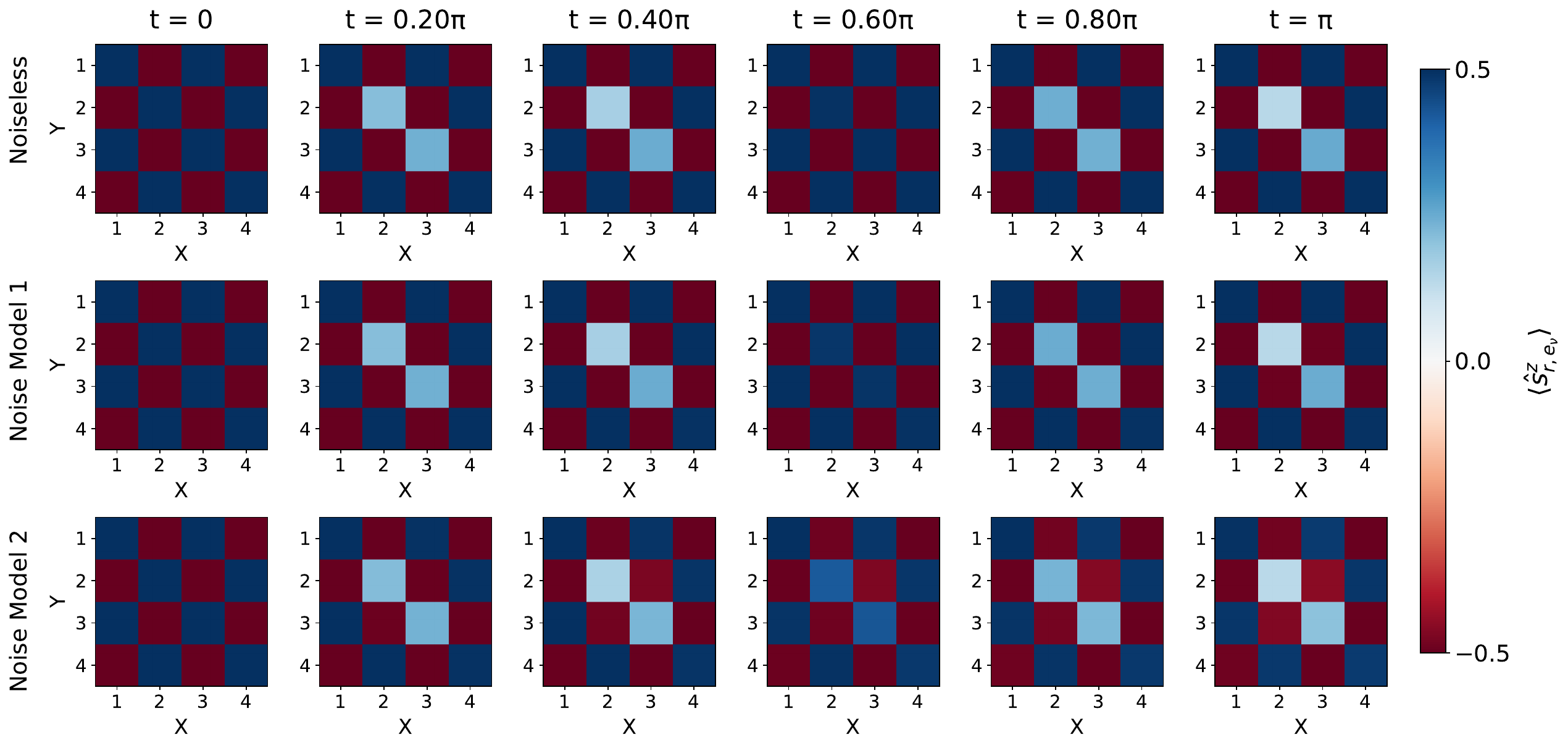}
    }

    \caption{Dynamics of $\mathcal{M}(t)$ over the four center spins in System II using parameters $m=0.42$, $\kappa = 1$ and $J=0$: (a) Comparison between noiseless Trotterized simulations and exact results using $d\theta = 0.01\pi$, (b) Comparison of noisy simulations against exact and noiseless results for $d\theta = 0.05\pi$, (c) Snapshots of the lattice at representative times, highlighting the local $\langle \hat{s}^z_{\mathbf{r, e_{\nu}}} \rangle$  dynamics under different noise conditions relative to the noiseless case.}
    \label{fig:vacuum_magnetization_combined}
\end{figure*}

\section{Numerical Results}
\label{Sec:NumericalResults}
We employ the \texttt{MQT-Qudits} package~\cite{Mato2022, Mato2023, Mato2023_2, Mato2023_3, Mato2024}, a \texttt{Python}-based framework developed for simulating qudit-based quantum circuits. To model the time evolution of the system, we implement a first-order Suzuki-Trotter decomposition \eqref{u_trotter} to approximate the system's time evolution. 

In order to investigate the robustness of our circuits under realistic experimental conditions, we introduce two distinct noise models:
\begin{itemize}
    \item Noise Model 1 -- Applies dephasing and depolarizing noise to all gate operations using error rates, where $p_{\text{1q}} = 3\times 10^{-7}$ for single qudit gates, $p_{\text{CX}} = 2 \times 10^{-5}$ and $p_{\text{MS}} = 1 \times 10^{-5}$ for entangling gates, with the same probability for each noise type.
    \item Noise Model 2 -- Assumes the same noise types as Model 1, but with rates that are an order of magnitude larger. While this is still better than the current state of the art, we believe such noise rates to be achievable in the near term.
\end{itemize} 

\begin{figure*}[t]
\captionsetup[subfloat]{position=top,justification=raggedright,singlelinecheck=false, font=large}
    \centering
    \subfloat[]{
    \hspace*{-2em}
        \includegraphics[width=0.5\textwidth]{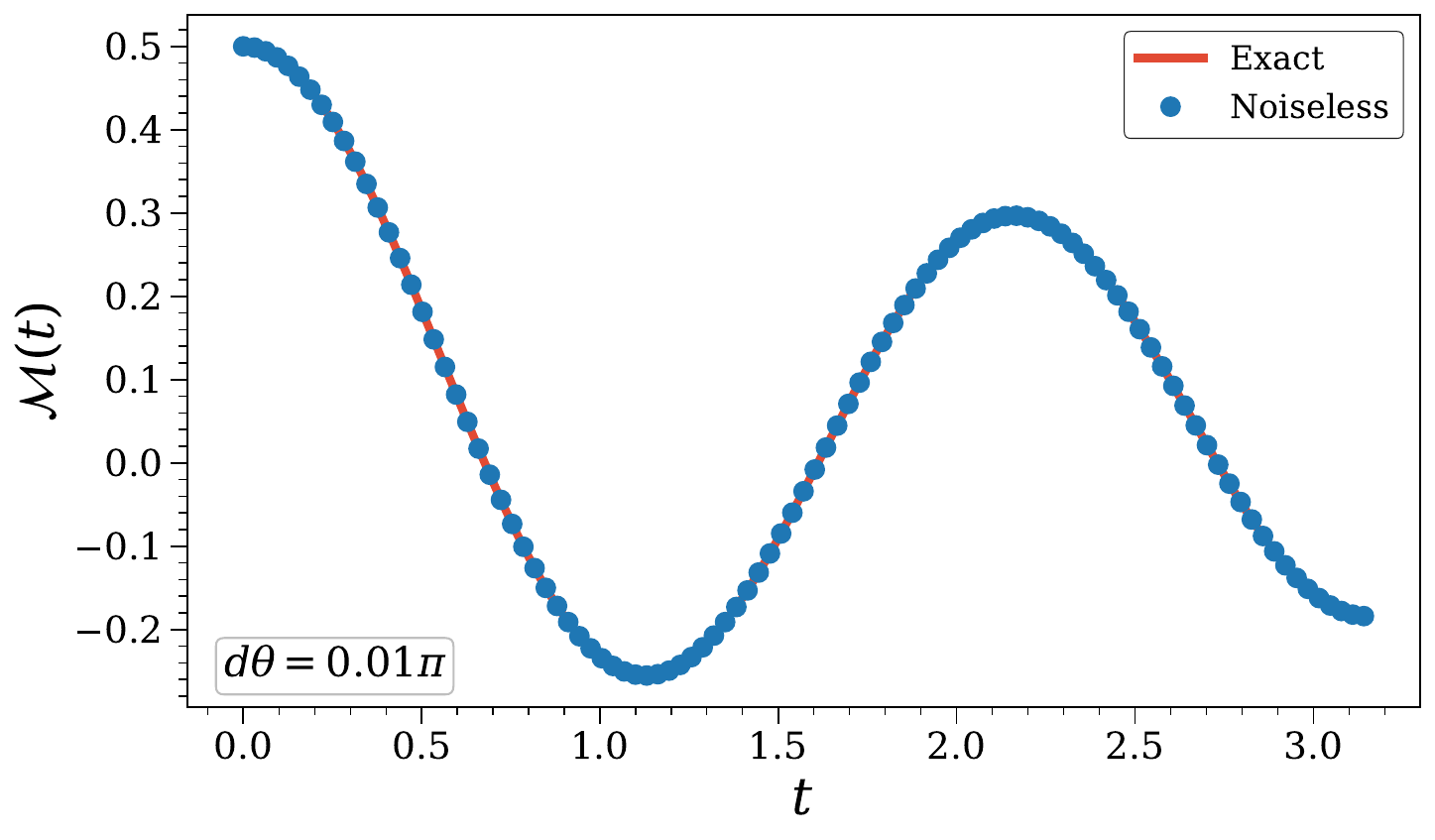}
    }
    \hfill
    \subfloat[]{
        \includegraphics[width=0.5\textwidth]{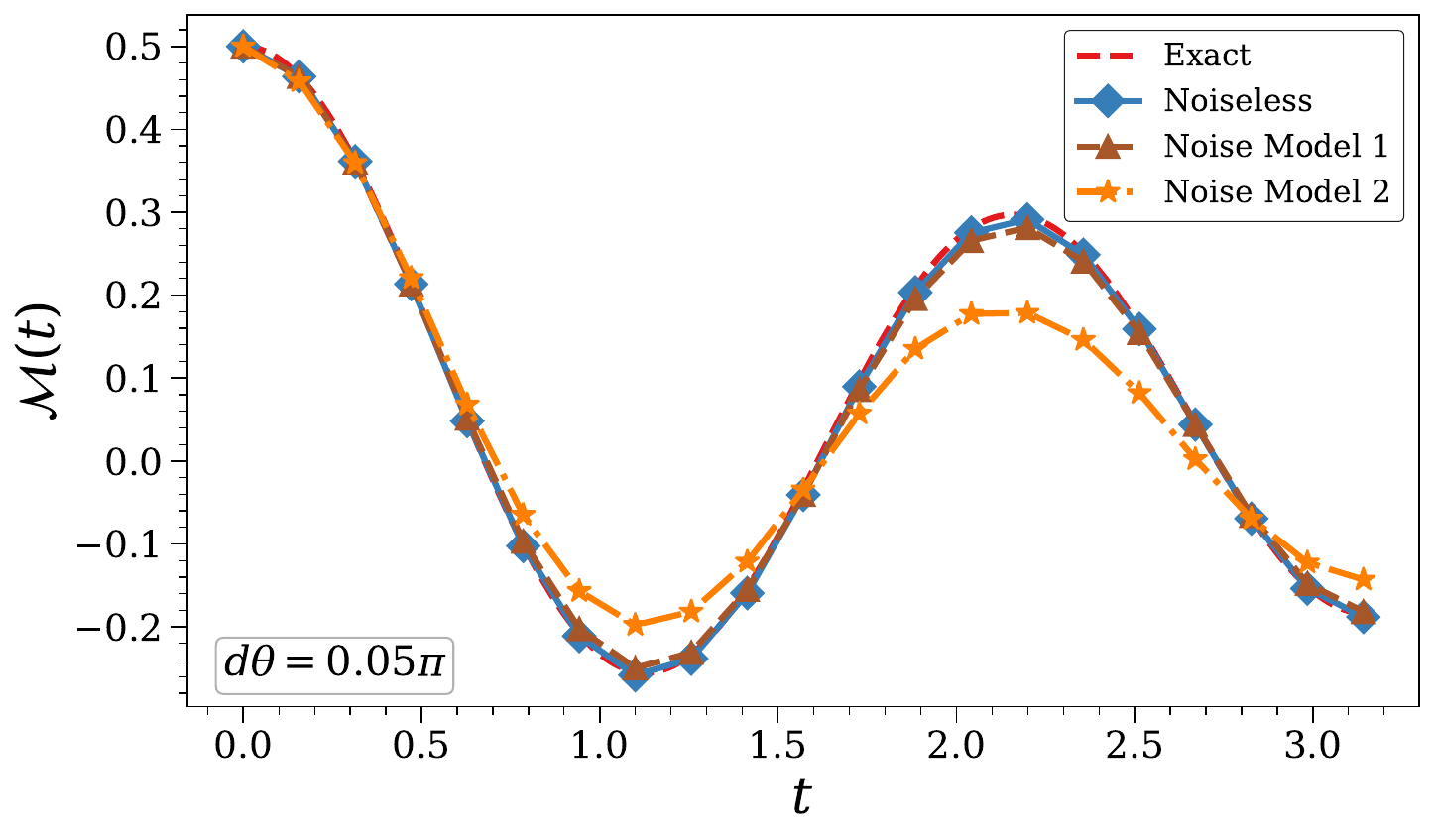}
    }\hfill


    \subfloat[]{
        \includegraphics[width=\textwidth]{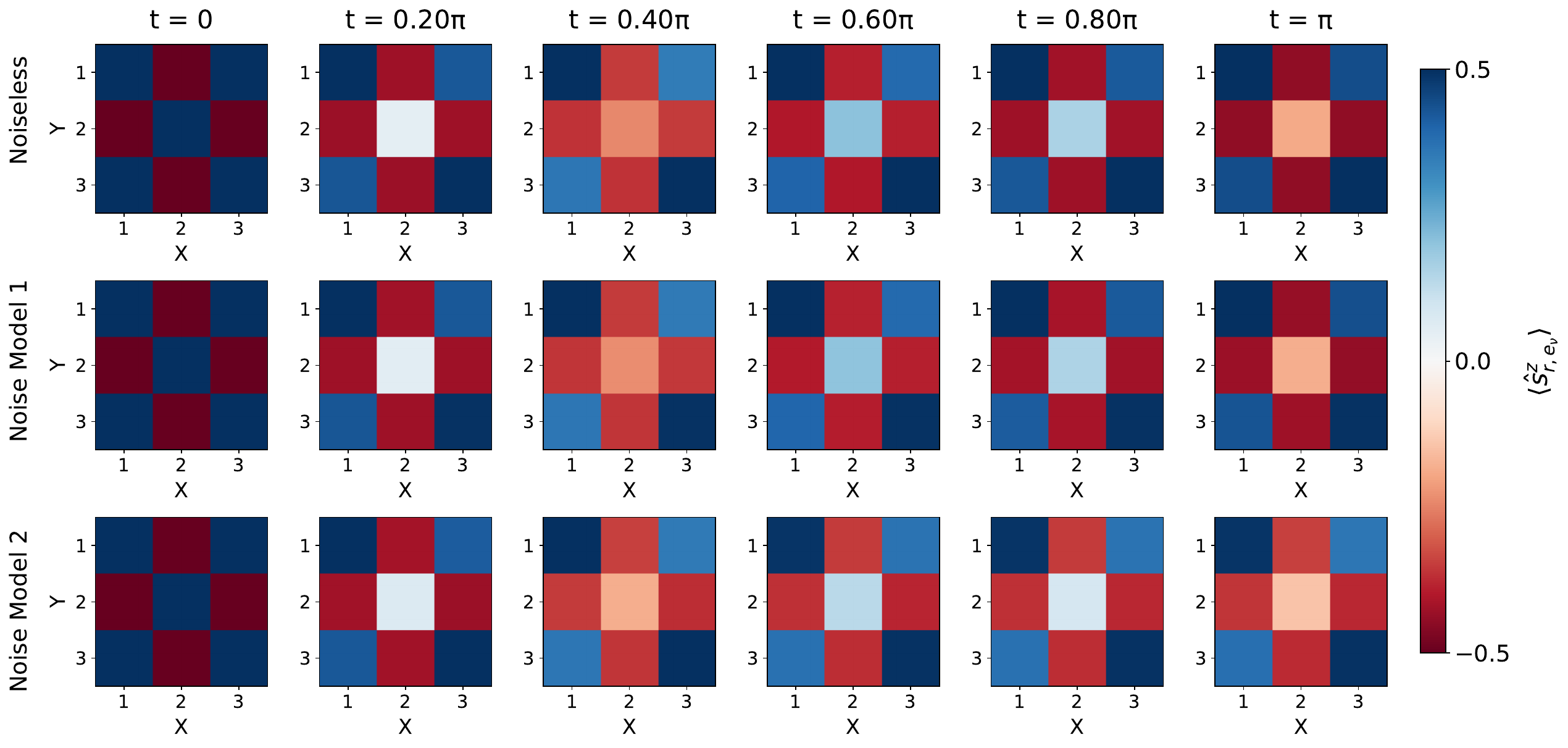}
    }

    \caption{Dynamics of $\mathcal{M}(t)$ for the center spin in System III using parameters $m=0.42$, $\kappa = 1$ and $J=0.5$: (a) Comparison between noiseless Trotterized simulations and exact results using $d\theta = 0.01\pi$, (b) Comparison of noisy simulations against exact and noiseless results for $d\theta = 0.05\pi$, (c) Snapshots of the lattice at representative times, highlighting the local $\langle \hat{s}^z_{\mathbf{r, e_{\nu}}} \rangle$ dynamics under different noise conditions relative to the noiseless case.}
    \label{fig:9_qudit_combined_magnetization}
\end{figure*}
As our primary observable, we analyze the dynamics of magnetization $\mathcal{M}(t)$, which is the averaged sum of the expectation values of local $\hat{s}^z_{\mathbf{r, e_{\nu}}}$ operators, defined as
\begin{equation}
\mathcal{M}(t) = \frac{1}{N} \sum_{\mathbf{r}, e_{\nu} \in A} \langle \psi(t) | \hat{s}_{\mathbf{r}, e_{\nu}}^z | \psi(t) \rangle,
\end{equation}
where $A$ and $N$ are the set and number of spins addressed, respectively.

Despite the scalability of our approach, we focus here on small lattice sizes --- specifically \(3 \times 3\) and \(4 \times 4\) grids --- as the Hilbert space size increases rapidly with system size due to the qudit encoding, rendering circuit simulations infeasible beyond this scale. As the number of sites grows, not only does the dimensionality of the Hilbert space increase exponentially, but the number of quantum gates required also rise, leading to a substantial computational overhead. Consequently, larger lattices are currently intractable for efficient circuit simulations, justifying our focus on smaller systems.

\subsection{$4\times4$ lattice}
We now present simulations for quench dynamics on a $4\times4$ lattice of links, starting from two distinct initial states, referred to as System I and System II, as depicted in Fig.~\ref{fig:lattice_scheme}. System I corresponds to a filled matter configuration in the original model, while System II represents the vacuum state with no initial matter excitations. 

For both systems, we study the dynamics of $\mathcal{M}(t)$ where $A$ is the set containing the four spins in the center, as they are the only spins that can contribute to the dynamics. We consider the case without plaquette interactions and use parameters $m = 0.42$ and $\kappa = 1$. Including the plaquette terms in a $4 \times 4$ lattice leads to an enormous Hilbert space size of $2^4 \times 4^{12}$. To keep the problem computationally manageable, we exclude the plaquette terms from our dynamical simulations in this case. In doing so, we reduce the effective Hilbert space size to $2^4 \times 3^8 \times 4^4$, which, while still large, is tractable. This simplification allows us to simulate the dynamics of the system.

For a small Trotter step size $d\theta = 0.01\pi$, the noiseless Trotterized results show excellent agreement with the exact evolution, as shown in Fig.~\ref{fig:combined_magnetization_dynamics_snapshots}(a) for System I, validating the accuracy of the circuit construction. For noisy simulations, we increase the Trotter step size to $d\theta=0.05\pi$ to reduce circuit depth, while still keeping Trotter errors small enough to reliably capture the system's dynamics. Fig.~\ref{fig:combined_magnetization_dynamics_snapshots}(b) compares noisy circuit simulations for System I under the two considered noise models. Under Noise Model 2, the dynamics is well captured up to intermediate times, while Noise Model 1 yields even closer agreement with the noiseless case. The snapshots in Fig.~\ref{fig:combined_magnetization_dynamics_snapshots}(c) demonstrate that local electric field dynamics in noisy simulations remain in good qualitative agreement with the noiseless results.

In System I, the dynamics is governed by three dynamical spins that contribute to the evolution. Fig.~\ref{fig:combined_magnetization_dynamics_snapshots} exhibits clear signatures of recurrent oscillations in $\mathcal{M}(t)$, arising from the simultaneous excitation and de-excitation of these three spins.

The circuit simulations for System II also demonstrate the same qualitative behavior as for System I, with noiseless results closely matching exact dynamics, as shown in Fig.~\ref{fig:vacuum_magnetization_combined}(a), and noisy simulations preserving key features across both noise models, as shown in Fig.~\ref{fig:vacuum_magnetization_combined}(b,c). A similar oscillatory behavior is observed for System II as for System I, where here two dynamical spins drive the dynamics. These spins undergo synchronized excitation and de-excitation, resulting in coherent, recurrent oscillations in a similar fashion as System I. This consistent performance across different initial states supports the robustness of our Trotterized circuit approach for simulating quench dynamics in the $2+1$D QLM.

\subsection{$3\times3$ lattice}
In this section, we present the results for quench dynamics for a $3\times3$ lattice, where the initial state is considered as depicted in System III shown in Fig.~\ref{fig:lattice_scheme}. We incorporate all the terms in \eqref{h_mio_s=1/2}, including the plaquette term, and set the parameters to $m = 0.42$, $\kappa = 1$, and $J = 0.5$. This implies that the Hamiltonian includes a single coupling term and two distinct plaquette interaction contributions. In this system, we study the dynamics of $\mathcal{M}(t)$ where $A$ is the set containing only the center spin, which is acted upon by both types of terms. The interplay between these coupling and plaquette interactions leads to $\mathcal{M}(t)$ exhibiting coherent, but nonsinusoidal oscillations in contrast to System I and II; see Fig.~\ref{fig:9_qudit_combined_magnetization}.

In the noiseless setting, we find that the dynamics is accurately captured by the Trotterized circuit, as shown in Fig.~\ref{fig:9_qudit_combined_magnetization}(a). We also perform simulations with the noise models described earlier. As with the 
$4\times4$ lattice results, we find that despite the added complexity from the plaquette interactions, the circuit continues to reproduce key dynamical features of $\mathcal{M}(t)$ evolution with very good accuracy, as shown in Fig.~\ref{fig:9_qudit_combined_magnetization}(b,c), highlighting the feasibility of implementing such dynamics on near-term qudit hardware.

\section{Generating coupling term circuits for higher spins}\label{Sec:Higher_spin_circ}

In this section, we present a general approach for constructing qudit-based circuits, focusing on the systematic mapping of relevant states to facilitate the implementation of projector verification subcircuits, and subsequently, the complete circuits for the coupling term. We show this through explicit circuit construction for the spin-$1$ QLM. We encode the spin-$1$ system into qudits with a maximum local Hilbert space dimension of $7$. Our notation assigns the three-link states as $ \ket{0} $ for $ s_z = -1 $, $ \ket{1} $ for $ s_z = 0 $, and $ \ket{2} $ for $ s_z = 1 $. The remaining qudit states are labeled as $\ket{3}$, $\ket{4}$, and so on.

\begin{figure*}[t]
\captionsetup[subfloat]{position=top,justification=raggedright,singlelinecheck=false, font=large}
  \centering
  \subfloat[]{%
    \resizebox{\linewidth}{!}{%
\begin{quantikz}[row sep={2.7cm,between origins}, font=\Huge]
    \lstick{$\ket{a}$} & \gate[3][3cm][1cm]{\text{P}} & \qw & \midstick[3,brackets=none]{\textbf{\Huge =}} & \ctrl{1} & \ctrl{1} & \ctrl{1} & \ctrl{1} & \ctrl{1} & &  & & & & & &\\
    \lstick{$\ket{b}$} & & \qw & & \gate[1][2cm][2cm]{\text{CX}_{1, 0\leftrightarrow 1}} & \gate[1][2cm][2cm]{\text{CX}_{1, 2\leftrightarrow 3}} & \gate[1][2cm][2cm]{\text{CX}_{0, 0\leftrightarrow 2}} & \gate[1][2cm][2cm]{\text{CX}_{2, 1\leftrightarrow 3}} & \gate[1][2cm][2cm]{\text{CX}_{2, 2\leftrightarrow 4}} & \ctrl{1} & \ctrl{1} &\ctrl{1} & \ctrl{1} & \ctrl{1} & \ctrl{1} & &\\
    \lstick{$\ket{c}$} & & \qw & & & & & & & \gate[1][2cm][2cm]{\text{CX}_{0, 1\leftrightarrow 3}} & \gate[1][2cm][2cm]{\text{CX}_{0, 0\leftrightarrow 4}} & \gate[1][2cm][2cm]{\text{CX}_{1, 2\leftrightarrow 3}} &\gate[1][2cm][2cm]{\text{CX}_{1, 1\leftrightarrow 4}} & \gate[1][2cm][2cm]{\text{CX}_{3, 0\leftrightarrow 3}} & \gate[1][2cm][2cm]{\text{CX}_{2, 2\leftrightarrow 4}} & &\\
\end{quantikz}
    }%
  }\hfill
  \subfloat[]{%
    \resizebox{\linewidth}{!}{%
    \begin{quantikz}[row sep={2.7cm,between origins}, font=\Huge]
    \lstick{$\ket{a}$} & \gate[2][4cm][1cm]{\text{CR$^{a,b}_\mathrm{x}$}(\theta)} & \qw & \midstick[2,brackets=none]{\textbf{\Huge =}} & & & & & \ctrl{1} &\ctrl{1} & & & \ctrl{1} &\ctrl{1} & & & &\\
    \lstick{$\ket{b}$} & & \qw & & \gate[1][2cm][2cm]{\text{H}^{3,5}} & \gate[1][2cm][2cm]{\text{H}^{4,6}} & \gate[1][2cm][2cm]{\text{Rz}^{3,5}\big(\frac{\theta}{2}\big)} & \gate[1][2cm][2cm]{\text{Rz}^{4,6}\big(\frac{\theta}{2}\big)} & \gate[1][2cm][2cm]{\text{CX}_{3, 3\leftrightarrow 5}} & \gate[1][2cm][2cm]{\text{CX}_{4, 4\leftrightarrow 6}} & \gate[1][2cm][2cm]{\text{Rz}^{3,5}\big(-\frac{\theta}{2}\big)} & \gate[1][2cm][2cm]{\text{Rz}^{4,6}\big(-\frac{\theta}{2}\big)} & \gate[1][2cm][2cm]{\text{CX}_{3, 3\leftrightarrow 5}} &\gate[1][2cm][2cm]{\text{CX}_{4, 4\leftrightarrow 6}} & & \gate[1][2cm][2cm]{\text{H}^{3,5}} & \gate[1][2cm][2cm]{\text{H}^{4,6}} &\\
\end{quantikz}
    }%
  }
  \hfill
  \subfloat[]{%
    \resizebox{\linewidth}{!}{%
        \begin{quantikz}[row sep={1.8cm,between origins}, font=\Huge]
    \lstick{$\ket{P_L^1}$} & \gate[3][2.5cm]{\text{P}} & & & & & & & & & & & & & \gate[3][2.5cm]{\text{P}^{\dag}} &\\
    \lstick{$\ket{P_L^2}$} & & & & & & & & & & & & & & &\\
    \lstick{$\ket{P_L^3}$} & & \ctrl{1} & \ctrl{1} & \ctrl{1} & \ctrl{1} & & & & \ctrl{1} & \ctrl{1} & \ctrl{1} & \ctrl{1} & & &\\
    \lstick{$\ket{\text{targ}}$} & & \gate[1][1cm][1.8cm]{\text{CX}_{3, 1\leftrightarrow 3}} & \gate[1][1cm][1.8cm]{\text{CX}_{3, 0 \leftrightarrow 5}} & \gate[1][1cm][1.8cm]{\text{CX}_{4, 1\leftrightarrow 4}} & \gate[1][1cm][1.8cm]{\text{CX}_{4, 2\leftrightarrow 6}} & & \gate[2][2cm]{\text{CR$^{P_R^3, \text{targ}}_\mathrm{x}$}(\theta)} & & \gate[1][1cm][1.8cm]{\text{CX}_{4, 2\leftrightarrow 6}} & \gate[1][1cm][1.8cm]{\text{CX}_{4, 1\leftrightarrow 4}} & \gate[1][1cm][1.8cm]{\text{CX}_{3, 0\leftrightarrow 5}} & \gate[1][1cm][1.8cm]{\text{CX}_{3, 1\leftrightarrow 3}} & & &\\
    \lstick{$\ket{P_R^3}$} & & & & & & \gate[3][2.5cm]{\text{P}} & & \gate[3][2.5cm]{\text{P}^{\dag}} & & & & & & &\\
    \lstick{$\ket{P_R^2}$} & & & & & & & & & & & & & & &\\
    \lstick{$\ket{P_R^1}$} & & & & & & & & & & & & & & &\\
    \end{quantikz}

}
}
    \caption{(a) The projector verification circuit $\mathrm{P}$ maps neighboring qudits according to the configurations specified in Tables~\ref{tab:s_1_p0} and~\ref{tab:s_1_p-1}. (b) The subcircuit $\mathrm{CR}^{ab}_\mathrm{x}$($\theta$) performs a controlled-X rotation by an angle $\theta$ within the appropriate subspaces, conditioned on the control qudit's state. The control qudit is labeled by index $a$ and the target qudit by $b$. $\mathrm{H}^{i,j}$ gate acts as the Hadamard gate in the $\{\ket{i}, \ket{j}\}$ subspace. (c) Full quantum circuit for implementing evolution operator corresponding to spin-$1$ $\hat{H}_C$ \eqref{h_qudit_qlm}:  The complete quantum circuit for the spin-$1$ coupling term includes 56 two-qudit gates and 8 single-qudit gates.
}
\label{fig:U_c_spin_1}
\end{figure*}
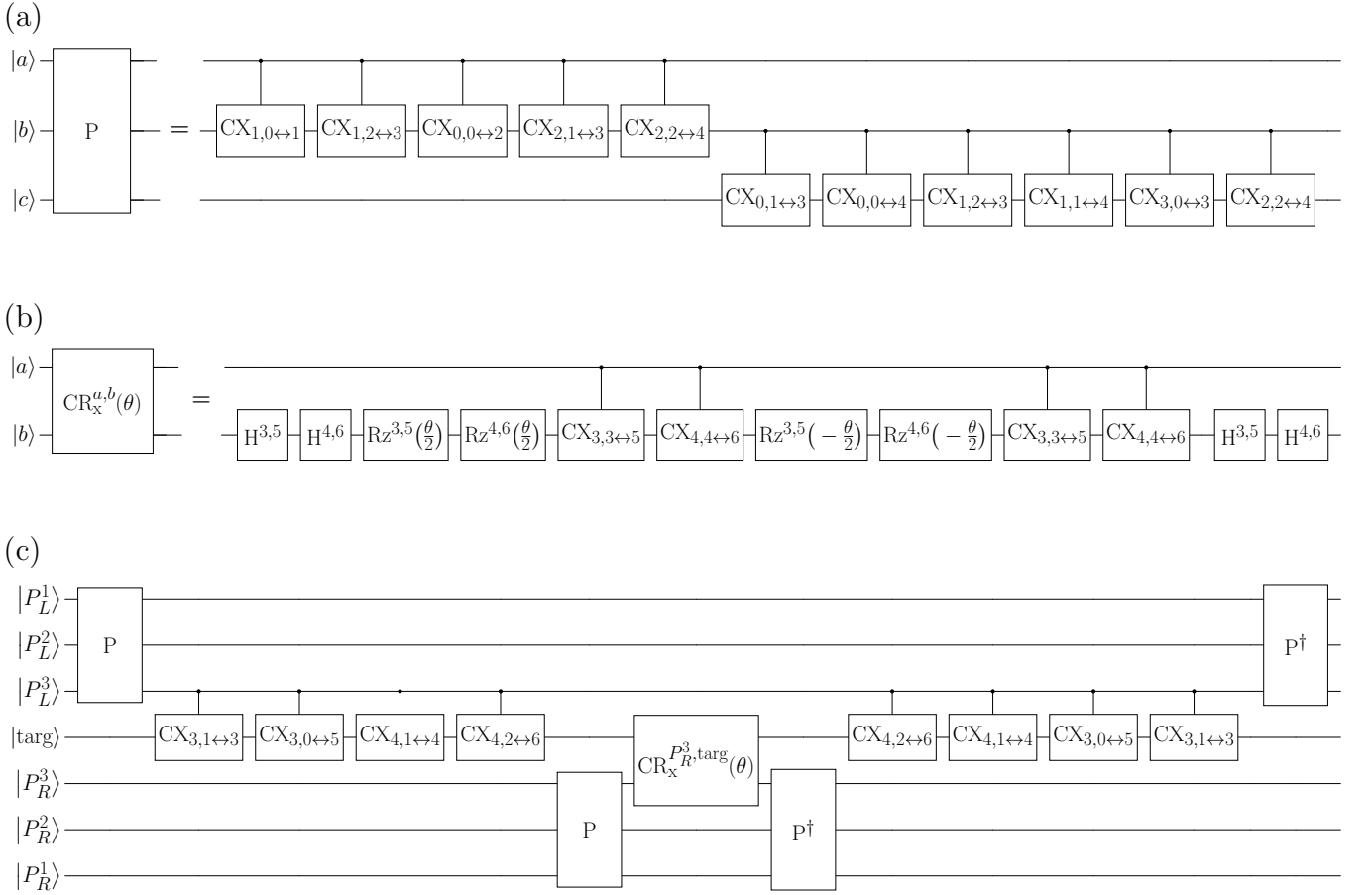

The circuit for implementing the unitary evolution operator $\hat{U}_{C_{\mathbf{r, e_{\nu}}}}^{S=1; \rm MIO}$ corresponding to the coupling term in $\hat{H}_{\mathrm{MIO}}^{S=1}$ \eqref{h_qudit_qlm} shown in Fig.~\ref{fig:U_c_spin_1}. To achieve this circuit decomposition, we use the identity \begin{equation}
e^{-\imath  \theta \hat{\sigma}^{x; a,b}_{\mathbf{r, e_{\nu}}}} = H^{a,b}_{\mathbf{r, e_{\nu}}} \, e^{-\imath  2\theta \hat{s}^{z;a,b}_{\mathbf{r, e_{\nu}}}}\, H^{a,b}_{\mathbf{r, e_{\nu}}},
\end{equation}
where $\{a, b\}$ denotes the target subspace. We also make use of the relations
\begin{equation}
\begin{aligned}
(\hat{P}^{0}_{\mathbf{r}} \hat{\sigma}^{x;12}_{\mathbf{r, r+e_{\nu}}} \hat{P}^{0}_{\mathbf{r, e_{\nu}}})^2 & = \hat{{I}}_{\hat{P}^0_{\mathbf{r}}} \hat{{I}}^{1,2}_{\mathbf{r, r+e_{\nu}}} \hat{{I}}_{\hat{P}^0_{\mathbf{r, e_{\nu}}}} =  \hat{{I}}',\\
(\hat{P}^{1}_{\mathbf{r}} \hat{\sigma}^{x;01}_{\mathbf{r, r+e_{\nu}}} \hat{P}^{1}_{\mathbf{r, e_{\nu}}})^2 & = \hat{{I}}_{\hat{P}^1_{\mathbf{r}}} \hat{{I}}^{0,1}_{\mathbf{r, r+e_{\nu}}}\hat{{I}}_{\hat{P}^1_{\mathbf{r, e_{\nu}}}} = \hat{{I}}'',
\end{aligned}
\end{equation}
to get 
\begin{equation}\label{unitary_coupling_s=1}
\begin{aligned}
\hat{U}_{C_{\mathbf{r, e_{\nu}}}}^{S=1; \rm MIO}(\theta) = & \sum_{n=0}^\infty\frac{\left(-\imath \theta \hat{P}^{0}_{\mathbf{r}} \hat{\sigma}^{x;12}_{\mathbf{r, r+e_{\nu}}} \hat{P}^{0}_{\mathbf{r, e_{\nu}}} + \hat{P}^{1}_{\mathbf{r}} \hat{\sigma}^{x;01}_{\mathbf{r, r+e_{\nu}}} \hat{P}^{1}_{\mathbf{r, e_{\nu}}}\right)^n}{n!}\\
= & \, \mathbb{1} - \hat{\mathrm{I}}' + \hat{P}^0_{\mathbf{r}}\, H^{12}_{\mathbf{r, e_{\nu}}} \, e^{-\imath 2\theta \hat{s}^{z;12}_{\mathbf{r, e_{\nu}}}}\, H^{12}_{\mathbf{r, e_{\nu}}} \, \hat{P}^0_{\mathbf{r+e_{\nu}}}\\
& + \hat{P}^1_{\mathbf{r}}\, H^{01}_{\mathbf{r, e_{\nu}}} \, e^{-\imath 2\theta \hat{s}^{z;01}_{\mathbf{r, e_{\nu}}}}\, H^{01}_{\mathbf{r, e_{\nu}}} \, \hat{P}^1_{\mathbf{r+e_{\nu}}}.
\end{aligned}
\end{equation}

We follow the same approach as employed for the spin-$1/2$ case to construct the circuit. First, we map the three-link states onto three-qudit states. Tables~\ref{tab:s_1_p0} and \ref{tab:s_1_p-1} present the mappings for all the three-link states that satisfy the projector conditions
\begin{equation}
\hat{P}_{\mathbf{r}}^{m} = \ketbra{\sum_{a \in L} s_a^z = m}_{\mathbf{r}}, \, m \in \{0, -1\}.
\end{equation}
Since there are two possible projectors for $S=1$, we ensure that for $\hat{P}_{\mathbf{r}}^{0}$, the last qudit is mapped to $ \ket{3} $, while for $\hat{P}_{\mathbf{r}}^{-1}$, it is mapped to $ \ket{4} $. This mapping strategy optimizes the circuit construction, requiring only 11 CX gates.

\begin{table}[H]
    \centering
    \begin{tabular}{|c|c|}
        \hline
        \textbf{3-link state} & \textbf{3-qudit state} \\  
        \hline
        $\ket{111}$ & $\ket{103}$ \\  
        \hline
        $\ket{120}$ & $\ket{123}$ \\  
        \hline
        $\ket{102}$ & $\ket{113}$ \\  
        \hline
        $\ket{021}$ & $\ket{003}$ \\  
        \hline
        $\ket{012}$ & $\ket{013}$ \\  
        \hline
        $\ket{210}$ & $\ket{233}$ \\  
        \hline
        $\ket{201}$ & $\ket{203}$ \\  
        \hline
        
    \end{tabular}
    \caption{Mapping 3-link states that satisfy $ \hat{P}_{\mathbf{r}}^{0}$ condition for spin-$1$ QLM}
    \label{tab:s_1_p0}
\end{table}

\begin{table}[H]
    \centering
    \begin{tabular}{|c|c|}
        \hline
        \textbf{3-link state} & \textbf{3-qudit state} \\  
        \hline
        $\ket{110}$ & $\ket{104}$ \\  
        \hline
        $\ket{101}$ & $\ket{114}$ \\  
        \hline
        $\ket{011}$ & $\ket{014}$ \\  
        \hline
        $\ket{002}$ & $\ket{024}$ \\  
        \hline
        $\ket{020}$ & $\ket{004}$ \\  
        \hline
        $\ket{200}$ & $\ket{204}$ \\  
        \hline  
    \end{tabular}
    \caption{Mapping 3-link states that satisfy $ \hat{P}_{\mathbf{r}}^{-1}$ condition for spin-$1$ QLM}
    \label{tab:s_1_p-1}
\end{table}

To construct the complete circuit for the coupling term in the spin-$1$ QLM, we first apply the projector verification circuit to the left neighbors. This determines the appropriate controlled transitions to be performed on the target qudit based on the satisfied projector condition.
Subsequently, the same projector verification circuit is applied to the right neighbors, where the controlled rotation is executed within the designated subspace afterwards. Finally, the sequence of operations is reversed to restore all qudits to the $\{\ket{0},\ket{1},\ket{2}\}$ subspace. First, the projector verification circuit for the right neighbors is reversed, followed by the reversal of the controlled transitions applied due to the left neighbors. Lastly, the projector verification circuit for the left neighbors is reversed, ensuring that all qudits return to the original subspace. This approach efficiently implements the coupling term using a total of 56 gates for the spin-$1$ QLM; cf.~Fig.~\ref{fig:U_c_spin_1}. The same approach can be extended to construct circuits for the coupling term for higher-spin systems. 

While we presented the complete circuit construction for the coupling term in the spin-$1$ QLM here, we do not include simulations, as scaling to system sizes large enough to capture nontrivial dynamics is challenging. Since the circuits we propose utilize qudits with a maximal dimension of 7, they are, in principle, compatible with current qudit-based quantum processors \cite{Ringbauer_2022}. This compatibility opens up the possibility of experimental realization, potentially enabling simulations beyond the reach of classical methods.

\section{Conclusions}
\label{Conclusions}
In this work, we have presented an efficient and scalable framework for the digital quantum simulation of $2+1$D U(1) QLMs with dynamical matter using qudit-based architectures. By integrating out the matter fields analytically, we reduced the model to the Hamiltonian \eqref{h_qudit_qlm}, containing only gauge degrees of freedom. This formulation naturally maps onto qudit architectures, allowing each gauge degree of freedom to be represented by a single qudit, eliminating the need for decomposition into multiple qubits and vastly reducing circuit resources.

Building on this approach, we constructed explicit quantum circuits for the full Hamiltonian in the spin-$1/2$ case \eqref{h_mio_s=1/2}. For $S=1/2$, each link per Trotter step requires only $22$ entangling gates and $4$ single-qudit gates. The plaquette interaction, which involves four-body terms, is implemented using $38$ entangling gates per Trotter step. Our design for the minimal coupling term is generalizable to higher-spin systems and relies on efficient decomposition into projector subcircuits. The generalization to higher spins is also laid out with an explicit circuit construction for the spin-$1$ representation. 

Furthermore, we also showed through numerical simulations that the exact quench dynamics can be accurately reproduced by our circuits, even under realistic noise conditions with experimentally achievable error rates. Overall, our results showcase the feasibility of near-term quantum simulators  in probing nontrivial $2+1$D gauge-theory dynamics with high fidelity. This qudit-based implementation, grounded in the matter-integrated-out formulation, offers significant advantages over qubit-only schemes by reducing circuit depth, gate count, and the overall number of qubits required --- since the matter fields are analytically eliminated. As a result, it provides a resource-efficient framework ideally suited for probing real-time phenomena in $2+1$D LGTs on qudit-based quantum processors.

Looking forward, the methods developed here offer multiple directions for future research. One natural extension is to explore other lattice gauge theories --- both Abelian and non-Abelian --- particularly those where matter can be integrated out in a similar fashion, yielding Hamiltonians ideally suited for qudit-based implementations. The resource efficiency outlined by our construction here enables large-scale implementations on near-term qudit hardware, where a suite of phenomena relevant to both HEP and condensed matter can be investigated. The scalability offered by our approach is a step forward towards achieving quantum advantage in HEP quantum simulators. Indeed, applying advanced error mitigation strategies to these circuits could be a research direction to further enhance the accuracy of quantum simulations. As quantum hardware continues to improve, combining efficient qudit circuit designs with experimental error suppression techniques may enable probing the real-time dynamics of lattice gauge theories into competitively large evolution times in higher spatial dimensions, an ideal venue for quantum advantage.

\footnotesize{\begin{acknowledgments}
    The authors acknowledge stimulating discussions with Debasish Banerjee, Philipp Hauke, Zlatko Papić, Arnab Sen, and Bing Yang. R.J., J.C.L., J.J.O., and J.C.H.~acknowledge funding by the Max Planck Society, the Deutsche Forschungsgemeinschaft (DFG, German Research Foundation) under Germany’s Excellence Strategy – EXC-2111 – 390814868, and the European Research Council (ERC) under the European Union’s Horizon Europe research and innovation program (Grant Agreement No.~101165667)—ERC Starting Grant QuSiGauge. M.R., and M.M. acknowledge funding by the European Union under the Horizon Europe Programme---Grant Agreements 101080086---NeQST and by the European Research Council (ERC, QUDITS, 101039522). Views and opinions expressed are however those of the author(s) only and do not necessarily reflect those of the European Union or the European Research Council Executive Agency. Neither the European Union nor the granting authority can be held responsible for them. We also acknowledge support by the Austrian Science Fund (FWF) through the EU-QUANTERA project TNiSQ (N-6001), by the IQI GmbH, and by the Austrian Federal Ministry of Education, Science and Research via the Austrian Research Promotion Agency (FFG) through the project FO999914030 (MUSIQ) funded by the European Union-NextGenerationEU.
    This work is part of the Quantum Computing for High-Energy Physics (QC4HEP) working group.
\end{acknowledgments}}
\normalsize

\bibliography{biblio}

\end{document}